%% file: 1Dchain.tex
\documentclass[letter,11pt]{article}
\pdfoutput=1
\usepackage[textwidth=16cm,textheight=24cm]{geometry}
\usepackage[pdftex]{graphicx,color}
\usepackage{subfigure}
\usepackage{amsmath}
\usepackage{amsfonts}
\usepackage{amssymb}

\providecommand{\U}[1]{\protect\rule{.1in}{.1in}}

\ifx.epstexversion\undefined
\else
\fi
\expandafter\ifx\csname package@font\endcsname\relax\else
\expandafter\expandafter
\expandafter
\expandafter
\expandafter{\csname package@font\endcsname}
\fi

\newfont{\bbold}{msbm9 scaled \magstep1}
\newcommand{\bra}[1]{\langle #1|}
\newcommand{\ket}[1]{|#1\rangle}

\newcommand{\beq}{\begin{eqnarray}}
\newcommand{\eeq}{\end{eqnarray}}

\newcommand{\w}{\overline{w}}
\newcommand{\z}{\overline{z}}

\newcommand{\ba}{\begin{array}{ccc}}
\newcommand{\ea}{\end{array}}
\newcommand{\nn}{\nonumber \\}

\title{Detailed analysis of the continuum limit of a\\
supersymmetric lattice model in 1D}

\author{L. Huijse}
\begin{document}
\maketitle

\begin{center}
\small{Department of Physics, Harvard University, Cambridge MA 02138, USA\\
Huijse@physics.harvard.edu}
\end{center}

\begin{abstract}
We present a full identification of lattice model properties with their field theoretical counter parts in the continuum limit for a supersymmetric model for itinerant spinless fermions on a one dimensional chain. The continuum limit of this model is described by an $\mathcal{N}=(2,2)$ superconformal field theory (SCFT) with central charge $c=1$. We identify states and operators in the lattice model with fields in the SCFT and we relate boundary conditions on the lattice to sectors in the field theory. We use the dictionary we develop in this paper, to give a pedagogical explanation of a powerful tool to study supersymmetric models based on spectral flow \cite{Huijse08b}. Finally, we employ the developed machinery to explain numerically observed properties of the particle density on the open chain presented in \cite{Beccaria05}.
\end{abstract}

\newpage

\tableofcontents

\section{Introduction}
In the past decade a model that incorporates supersymmetry in lattice models of strongly interacting spinless fermions has been introduced \cite{fendley-2003-90} and explored for a variety of lattices in one and higher dimensions \cite{fendley-2003-36,fendley-2005-95,Fendley05,vanEerten05,Beccaria05,Jonsson06,Jonsson05p,Huijse08b,Huijse10, Huijse10b,Fendley10a,Fendley10b} (for reviews see \cite{Huijse08a,HuijseT10}). These models show various interesting features that are closely linked to the supersymmetry. First, supersymmetry provides a rich mathematical structure that allows for a considerable degree of analytic control in the regime where more standard perturbative techniques would fail. Second, it induces a delicate balance between kinetic and potential terms, which results in a strong form of quantum charge frustration. This so-called 'superfrustration', which typically occurs for two (or higher) dimensional lattices, is characterized by an extensive ground state entropy \cite{fendley-2005-95,vanEerten05}. For the one dimensional chain, the model was found to be integrable and quantum critical. In recent work \cite{Fendley10a,Fendley10b}, it was shown that the one dimensional model also enjoys special features, such as scale-free properties, which are again directly related to supersymmetry.

In this paper, we focus on the supersymmetric model on the one dimensional chain. A Bethe Ansatz soltution for this model was presented in \cite{fendley-2003-90,fendley-2003-36}. The continuum limit is described by an $\mathcal{N}=(2,2)$ superconformal field theory (SCFT) with central charge $c=1$. Here, we present a full identification of lattice model properties with their field theoretical counter parts in the continuum limit. In particular, we identify states and operators in the lattice model with fields in the SCFT and we relate boundary conditions on the lattice to sectors in the field theory. We will see that this model forms a textbook example of how a superconformal field theory can be identified studying the finite size lattice model properties. Apart from finite size scaling of the spectrum, we will study a boundary twist, entanglement entropy and one-point functions. The boundary twist in the lattice model is related to a spectral flow in the continuum theory. The power of this type of analysis was demonstrated in \cite{Huijse08b} where it was used to identify quantum criticality in various ladder models. The discussion here serves as a pedagogical introduction to this technique. Finally, the dictionary developed in this paper allows one to study various properties of the model. As an example we compute the density of the chain with open boundary conditions. This property was studied in \cite{Beccaria05}, where its scaling dimension was identified numerically. Furthermore, a remarkable $\mathbb{Z}_3$ substructure was observed. The result we obtain here, both confirms and explains these observations.

The paper is organised as follows. We first introduce the model and discuss some of the basics of supersymmetry. We then discuss the $\mathcal{N}=(2,2)$ superconformal field theory with central charge $c=1$. In section \ref{sec:lattice}, we present the full identification of this theory with the lattice model with closed boundary conditions. The next section provides a detailed explanation of the spectral flow analysis. Section \ref{sec:entangS} briefly presents some results for the entanglement entropy. In section \ref{sec:openbc}, we extend the dictionary to the case of open boundary conditions, which is put to work in section \ref{sec:obc2}, where the site-dependend particle density is computed on the field theory side.

\section{The model}\label{sec:model}
In quantum mechanics, supersymmetric theories are characterized by a positive 
definite energy spectrum and a twofold degeneracy of
each non-zero energy level. The two states with the same energy are called 
superpartners and are related by the nilpotent supercharge operator. 
Let us consider an $\mathcal{N}=2$ supersymmetric theory, defined by two
nilpotent supercharges $Q$ and $Q^{\dag}$ \cite{Witten82},
\begin{eqnarray}\nonumber
Q^2=(Q^{\dag})^2=0
\end{eqnarray}
and the Hamiltonian given by 
\begin{eqnarray}\nonumber
H=\{Q^{\dag},Q\} .
\end{eqnarray}
From this definition it follows directly that $H$ is positive definite:
\begin{eqnarray*}
\bra{\psi}H\ket{\psi}
  &=& \bra{\psi}(Q^{\dag}Q+QQ^{\dag})\ket{\psi}
\nonumber \\[2mm]
  &=&|Q\ket{\psi}|^2+|Q^{\dag}\ket{\psi}|^2 \geq 0 \ .
\end{eqnarray*}
Furthermore, both $Q$ and $Q^{\dag}$ commute with the Hamiltonian, 
which gives rise to the twofold degeneracy in the energy spectrum. 
In other words, all eigenstates with an energy $E_s>0$ form doublet
representations of the supersymmetry algebra. A doublet consists of two 
states $\ket{s}, Q \ket{s}$, such that $Q^{\dag}\ket{s}=0$. Finally, 
all states with zero energy must be singlets: 
$Q \ket{g}=Q^{\dag}\ket{g}=0$ and conversely, all singlets must be zero 
energy states \cite{Witten82}. In addition to supersymmetry our models also 
have a fermion-number symmetry generated by the operator $F$ with
\begin{eqnarray}\nonumber
[F,Q^{\dag}]=-Q^{\dag} \quad \textrm{and} \quad [F,Q]=Q.
\end{eqnarray}
Consequently, $F$ commutes with the Hamiltonian.

We now make things concrete and define a supersymmetric model for spin-less
fermions on a lattice, following
\cite{fendley-2003-90}. The operator 
that creates a fermion on site $i$ is written as $c_i^{\dag}$ with
$\{c_i^{\dag},c_j\}=\delta_{ij}$. To obtain a non-trivial Hamiltonian, we dress the fermion 
with a projection operator: 
$P_{<i>}=\prod_{j \textrm{ next to } i} (1-c_j^{\dag}c_j)$, which requires 
all sites adjacent to site $i$ to be empty. With $Q=\sum c_i^{\dag} P_{<i>}$ 
and $Q^{\dag}=\sum c_i P_{<i>}$, the Hamiltonian of these hard-core 
fermions reads
\begin{eqnarray}\label{Hsusygen}\nonumber
H=\{Q^{\dag},Q\}= 
\sum_i \sum_{j\textrm{ next to }i} P_{<i>} c_i^{\dag} c_j P_{<j>} 
       + \sum_i P_{<i>}.
\end{eqnarray}
The first term is just a nearest neighbor hopping term for hard-core fermions, 
the second term contains a next-nearest neighbor repulsion, a chemical 
potential and a constant. The details of the latter terms will depend on the 
lattice we choose. 

In this paper we consider the supersymmetric model on a one dimensional chain. For a chain of length $L$, the Hamiltonian takes on the explicit form:
\begin{eqnarray}
H=\sum_{i=1}^{L} \left[ P_{i-1} \big(c_i^{\dag} c_{i+1} + c_{i+1}^{\dag} c_i \big)  P_{i+2} \right] + \sum_{i=1}^{L} (n_i n_{i+2}) + L - 2 F .\nonumber
\end{eqnarray}
Here $P_i=1-n_i$, $n_i=c^{\dag}_i c_i$ is the usual number operator and $F=\sum_i n_i$ is the total number of fermions.

\section{Continuum theory}
The supersymmetric model on the chain can be solved exactly
through a Bethe Ansatz \cite{fendley-2003-90}. In the continuum
limit one can derive the thermodynamic Bethe Ansatz equations. The
model has the same thermodynamic equations as the XXZ Heisenberg
spin chain at a specific value of the anisotropy parameter
$\Delta$. There is indeed a mapping between the supersymmetric
model on the chain and the Heisenberg spin chain with special
boundary conditions \cite{fendley-2003-36}. The hamiltonian of the
XXZ chain is defined in terms of the usual Pauli matrices as
\begin{equation}\label{eq:XXZ}
 H_{\textrm{XXZ}} = \frac{1}{2} \sum_{i=1}^{L} \left[ \sigma^{x}_{i} \sigma^{x}_{i+1}+\sigma^{y}_{i} \sigma^{y}_{i+1} -\Delta \sigma^{z}_{i} \sigma^{z}_{i+1} \right].
\end{equation}
The continuum limit of the XXZ chain is described by the massless Thirring model \cite{Thacker81}, or equivalently a free massless boson $\Phi$ with action \cite{Friedan88}
\begin{equation}\label{eq:freeboson}
S=\frac{g}{4 \pi}\int \textrm{d} x\,\textrm{d} t\ \left[(\partial_t \Phi)^2 -
(\partial_x\Phi)^2\right].
\end{equation}
The coupling constant $g$ is related to the anisotropy parameter
$\Delta$ in the XXZ chain. On the conformal field theory side, the
coupling constant $g$ is related to the compactification radius
$R$ of the free boson theory via $g=2/R^2$. The free boson theory
is characterized by a central charge $c=1$ and a set of highest
weight states which depend on the compactification radius. At a
compactification radius $R=\sqrt{3}$, the conformal algebra is
enhanced to an $\mathcal{N}=(2,2)$ superconformal algebra
\cite{Waterson86,Friedan88}. The $(2,2)$ means that both the
holomorphic as well as the anti-holomorphic fields satisfy an
$\mathcal{N}=2$ superconformal algebra. More specifically, the
free boson at compactification radius $R=\sqrt{3}$ is the simplest
field theory with $\mathcal{N}=(2,2)$, namely the first in the
series of minimal supersymmetric models. It turns out that the
compactification radius $R=\sqrt{3}$ corresponds to an anisotropy
parameter of $\Delta=-1/2$ (see for example \cite{Affleck88})
which is precisely the value one obtains upon mapping the
supersymmetric model onto the XXZ chain \cite{fendley-2003-36}.

The fact that the low-energy spectrum of the supersymmetric model on the chain is described by a superconformal theory in the continuum limit, tells us that the model is quantum critical.

% \section{notes for me}
% In \cite{Friedan88} we have $\beta$. Here we have $g$ and then there is $R$. The relations are as follows:
% $g=\beta/2=2/R^2$. The dualities are $g \leftrightarrow 1/g$, $\beta \leftrightarrow 4/\beta$ (NB there is an error there in \cite{Friedan88}) and $R \leftrightarrow 2/R$. Finally, Ginsparg has $r=R/2$, with duality $r \leftrightarrow 1/2r$.

% \section{Free boson with $\mathcal{N}=(2,2)$ supersymmetry}\label{sec:freeboson}

\subsection{Superconformal algebra at $R=\sqrt{3}$}
In an $\mathcal{N}=2$ superconformal field theory \cite{Boucher86,DiVecchia85} there are three generators besides the stress-energy tensor: two supercharges, $G^+(z)$ and $G^-(z)$, with conformal dimension $3/2$ and a $U(1)$ current, $J(z)$, with conformal dimension 1. The $\mathcal{N}=2$ superconformal algebra is then given by the Virasoro algebra
\begin{eqnarray}\label{eq:virasoroalg}
 \left[ L_m,L_n \right] = (m-n)L_{m+n} + \frac{1}{12} c (m^3-m) \delta_{m+n,0}.
\end{eqnarray}
together with a $U(1)$ Kac-Moody algebra for the current
\begin{eqnarray}\label{eq:kacmoodyalg}
\left[ J_m,J_n \right]= \frac{c}{3} m \delta_{m+n,0} \quad \left[ L_m , J_n \right]=-n J_{m+n},
\end{eqnarray}
and the algebra of the supercharges
\begin{eqnarray}\label{eq:N=2comm1}
\left[ L_m, G^{\pm}_r \right] &=& (\frac{1}{2} m-r) G^{\pm}_{m+r}, \\
\left[ J_m, G^{\pm}_r \right] &=& \pm G^{\pm}_{m+r}, \label{eq:N=2comm2}\\
\{G^{\pm}_r,G^{\mp}_s\} &=& 2 L_{r+s} \pm (r-s) J_{r+s} + \frac{1}{3} c (r^2-\frac{1}{4}) \delta_{r+s,0}. \label{eq:N=2comm3}
\end{eqnarray}
Here $r$ runs over all values in $\mathbb{Z}+\alpha$, with $\alpha$ a real number which determines the branch cut properties of $G^{\pm}(z)$. For $\alpha=0$ the theory is said to be in the Ramond sector and for $\alpha=1/2$ it is said to be in the Neveu-Schwarz sector.

We will now identify the supercharges and the $U(1)$ current in the free boson spectrum at compactification radius $R=\sqrt{3}$.
The free boson field $\Phi$ can be decoupled into left and right movers: $\Phi=\Phi_L+\Phi_R$ and the dual is defined as $\tilde{\Phi}=g(\Phi_L-\Phi_R)$. The left moving field obeys the following OPEs
\begin{equation}\nonumber
 \Phi_L(z)\Phi_L(w) \sim -\frac{1}{2g} \ln (z-w), \quad \partial \Phi_L(z) \partial \Phi_L(w) \sim -\frac{1}{2g}\frac{1}{(z-w)^2},
\end{equation}
and similarly for the right moving field
\begin{equation}\nonumber
 \Phi_R(\overline{z})\Phi_R(\overline{w}) \sim -\frac{1}{2g} \ln (\overline{z}-\overline{w}), \quad \partial \Phi_R(\overline{z}) \partial \Phi_R(\overline{w}) \sim -\frac{1}{2g}\frac{1}{(\overline{z}-\overline{w})^2}.
\end{equation}
The operators
\begin{equation}\nonumber
 V_{m,n}= \, \colon \negthickspace \exp (\imath m \Phi + \imath n \tilde{\Phi})\colon ,
\end{equation}
are called vertex operators. Here the semicolons imply normal ordering, in the following we will drop this notation and tacitly assume that normal ordering is taken care of. The vertex operators are primary fields with conformal dimensions:
\begin{equation}\label{eq:confdimvertex}
h_{L,R}=(m \pm gn)^2/(4g) ,
\end{equation}
with $m \in \mathbb{Z}$ and $n \in \mathbb{Z}/2$. Note that we label holomorphic and anti-holomorphic dimensions with $L$ and $R$, for left and right movers, respectively. From (\ref{eq:confdimvertex}) we find that for $R=\sqrt{3}$, and thus $g=2/3$, the operators $V_{\pm 1, \pm 3/2}$ have conformal dimensions $(h_L,h_R)=(3/2,0)$ and the operators $V_{\pm 1, \mp 3/2}$ have conformal dimensions $(h_L,h_R)=(0,3/2)$. These four operators are the two left mover supercharges and two right mover supercharges. Finally, the $U(1)$ currents of dimensions $(1,0)$ and $(0,1)$ are proportional to $\partial \Phi_L$ and $\partial \Phi_R$ respectively. The proportionality factor follows from comparing the OPEs of $\partial \Phi_{L,R}$ with the OPE of the U(1) current
\begin{eqnarray}\label{eq:U1current}
J(z)J(w) \sim \frac{c/3}{(z-w)^2} \Rightarrow J_{L,R}(z) = \pm \imath \sqrt{2g c/3} \partial \Phi_{L,R} = \pm \imath 2/3 \partial \Phi_{L,R} .
\end{eqnarray}
So these operators, together with the stress-energy tensor form an $\mathcal{N}=(2,2)$ superconformal algebra. That is, both the left-moving and right-moving operators generate an $\mathcal{N}=2$ superconformal algebra.

\subsection{Spectrum}\label{sec:spectrum}
The hamiltonian, i.e. the energy operator, is the generator of translations in the time direction. On the cylinder it reads
\begin{eqnarray}
 H=L_{L,0}+L_{R,0}-c/12.
\end{eqnarray}
It follows that for $c=1$ the energy of a state is given by $E=h_R+h_L-1/12$. Just as for the lattice model, we find that, if we define the two supercharges
\begin{eqnarray}
 G= \frac{1}{\sqrt{2}} (G^+_{L,0} - G^+_{R,0}) \quad G^{\dag}= \frac{1}{\sqrt{2}} (G^-_{L,0} - G^-_{R,0}),
\end{eqnarray}
we can write the hamiltonian on the cylinder as
\begin{eqnarray}
 H=L_{L,0}+L_{R,0} - \frac{c}{12}=\{G,G^{\dag}\}.
\end{eqnarray}
Note that these supercharges are defined in the Ramond sector.
Indeed it turns out that the supersymmetric structure as we
discussed it in section \ref{sec:model} is only fully realized in
the Ramond sector of the superconformal algebra. From the
commutator of the supercharges with the Virasoro generators
(\ref{eq:N=2comm1}), one easily verifies that the hamiltonian
commutes with $G$ and $G^{\dag}$. Furthermore, from the
commutation relation with the U(1) current, we find that the
hamiltonian also commutes with $F=J_{L,0}-J_{R,0}$. The supercharges satisfy
\begin{eqnarray}
 \left[ F, G \right] = - G \quad \left[ F, G^{\dag} \right] = G^{\dag}.
\end{eqnarray}
These are precisely the relations we found for the supersymmetric model. It thus follows that the spectrum of the hamiltonian in the Ramond sector will be positive definite and decomposes into zero-energy singlets and positive energy doublets.

\subsection{Supercharges and the $U(1)$ current}\label{sec:supercharges}
To see the action of the supercharges
\begin{eqnarray}\label{eq:supercharges}
G^{\pm}_L=V_{\pm 1,\pm 3/2} \quad \textrm{and} \quad G^{\pm}_R=V_{\pm 1,\mp 3/2}.
\end{eqnarray}
we consider the action of the supercharge $G^+_L$ on a state $V_{m,n}\ket{0}$.
The mode expansion for $G^+_L$ is given by
\begin{eqnarray}\nonumber
G^+_{L,l} = \oint \frac{dz}{2 \pi \imath} z^{l+1/2} G^+_{L} =\oint \frac{dz}{2 \pi \imath} z^{l+1/2}
V_{1,3/2} ,
\end{eqnarray}
where $l \in \mathbb{Z}$ in the Ramond sector and $l \in (\mathbb{Z}+1/2)$ in the NS sector.
Consequently, we have
\begin{eqnarray}\nonumber
G^+_{L,l}V_{m,n}\ket{0} &=& \oint \frac{dz}{2 \pi \imath} z^{l+1/2}V_{1,3/2}
V_{m,n}\ket{0} \nonumber\\
&=& \oint \frac{dz}{2 \pi \imath} z^{3/2m+n+l+1/2} V_{m+1,n+3/2}\ket{0},
\end{eqnarray}
where, in the second line, we used the OPE for $V_{1,3/2} (z) V_{m,n} (w,\overline{w})$:
\begin{eqnarray}\nonumber
V_{1,3/2} (z) V_{m,n} (w,\overline{w}) &\sim& (z-w)^{3/2m+n}  V_{m+1,n+3/2} (w,\overline{w}).
\end{eqnarray}

Similarly, we obtain
\begin{eqnarray}
 G^-_{L,l} V_{m,n} (0,0) \ket{0} &=& \oint \frac{dz}{2 \pi \imath} z^{-3/2m-n+l+1/2} V_{m-1,n-3/2}\ket{0}, \nonumber\\
 G^+_{R,l} V_{m,n} (0,0) \ket{0} &=& \oint \frac{dz}{2 \pi \imath} \overline{z}^{3/2m+n+l+1/2} V_{m+1,n-3/2}\ket{0}, \nonumber\\
 G^-_{R,l} V_{m,n} (0,0) \ket{0} &=& \oint \frac{dz}{2 \pi \imath} \overline{z}^{-3/2m-n+l+1/2} V_{m-1,n+3/2}\ket{0}. \nonumber
\end{eqnarray}
For the contour integrals to be well-defined the power of $z$ or $\overline{z}$ has to be integer. Since $l$ is integer (half-integer) in the Ramond (NS) sector, we find $3/2m+n$ is half-integer (integer) in the Ramond (NS) sector. This condition can be reformulated by saying that $(-1)^{m+2n}$ is $1$ in the Neveu-Schwarz sector and $-1$ in the Ramond sector.

Furthermore, we see that the first mode $G^+_{L,l}$ of $G^+_{L}$ that gives a non-zero contribution must obey $3/2m+n+l+1/2 \leq -1$, i.e. $l \leq -3/2 -3/2m-n$. Equivalently, we find for $G^-_{L,l}$: $l \leq -3/2 +3/2m+n$ and for $G^{\pm}_{R,l}$: $l \leq -3/2 \mp 3/2m \pm n$.

Let us now determine the $U(1)$ charges of the vertex operators. The left moving $U(1)$ charge, $q_L$, is defined by the OPE of the $U(1)$ current with a primary field $\psi$
\begin{eqnarray}
J_L(z)\psi(w) \sim \frac{q_L}{(z-w)}\psi(w) \nonumber
\end{eqnarray}
and similarly for $q_R$.
Using the OPE
\begin{eqnarray}
\partial \Phi_L (z) V_{m,n} (w,\overline{w}) &\sim& - \frac{\imath (3m+2n)}{4} \frac{V_{m,n}}{(z-w)} \nonumber
\end{eqnarray}
and a similar expression for the right movers, we find the $U(1)$ charges corresponding to the vertex operator $V_{m,n}$ to be
\begin{eqnarray}\label{eq:U1charges}
 q_{L,R}=n/3\pm m/2.
\end{eqnarray}

\subsection{Highest weight states and descendants}\label{sec:freebosonspectrum}
Let us identify the highest weight states. Note that the supercharges (\ref{eq:supercharges}) change $m$ by $\pm 1$, while leaving $(-1)^{m+2n}$ unchanged. Furthermore, by acting on a state with certain combinations of the supercharges, one can raise or lower $n$ by multiples of 3 while keeping $m$ fixed. From this, it follows that we need only three highest weight states per sector, since all other states can be generated from these states with the supercharges. For example, one can easily check that $ V_{0,-5/2} \ket{0} = G^+_{R,-1}G^-_{L,-1}V_{0,1/2} \ket{0}$ and $ V_{-1,-1} \ket{0} = G^-_{L,-1}V_{0,1/2} \ket{0}$.

In the Ramond sector, we choose the highest weight states $V_{0,\pm 1/2} \ket{0}$ and $V_{0,3/2} \ket{0}$, since these are the states with lowest energy. Their conformal dimensions are $h_{L,R}=1/24$ and $h_{L,R}=3/8$, respectively, and thus their energies are $E=0$ and $E=2/3$. Clearly, the same reasoning applies in the Neveu-Schwarz sector and the highest weight states are $V_{0,0}\ket{0}$ and $V_{0,\pm 1}\ket{0}$ with $h_{L,R}=0$ and $h_{L,R}=1/6$ respectively. Note that these are precisely the conformal dimensions of the first minimal model in the supersymmetric minimal series \cite{Feigin82,Feigin85}. Using (\ref{eq:U1charges}), one also readily verifies the agreement with the first minimal model for the $U(1)$ charges of the highest weight states.

The spectrum is generated from the highest weight states by acting with the Virasoro generators and the supercharges. The action of the Virasoro generators on a highest weight state $\ket{h_L,h_R}$ is well known
\begin{eqnarray}
L_{L/R,0} \ket{h_L,h_R} &=& h_{L,R} \ket{h_L,h_R},\nonumber\\
L_{L/R,n} \ket{h_L,h_R} &=& 0, \nonumber\\
L_{L,-n} \ket{h_L,h_R} &=& \ket{h_L+n,h_R}, \nonumber\\
L_{R,-n} \ket{h_L,h_R} &=& \ket{h_L,h_R+n}, \nonumber
\end{eqnarray}
with $n>0$. The vacuum is defined as the state with $L_{L/R,0}\ket{0}=0$.

Supersymmetry implies that the zero energy states in the Ramond sector do not have superpartners. Consequently, they must be annihilated by the zero modes of the supercharges, that is $G^{\pm}_{L/R,0}$. From the inequalities relating the mode $l$ to $m$ and $n$ given in section \ref{sec:supercharges}, we find that indeed  $G^{\pm}_{L/R,0}V_{0,\pm 1/2} \ket{0}=0$. The third highest weight state in the Ramond sector, $h_{L,R}=3/8$, has non-zero energy, so we would expect this state to have a superpartner. In fact, since the left and right movers completely decouple in the continuum limit, the continuum theory has two $\mathcal{N}=2$ supersymmetries. Consequently, the third highest weight state forms a quadruplet instead of a doublet. We find that there are four states with $h_{L,R}=3/8$ and energy $E=3/8+3/8-1/12=2/3$, which are all related via the supercharges
\begin{eqnarray}
G^-_{L,0} V_{0,3/2} \ket{0} &=& V_{-1,0} \ket{0}, \nonumber\\
G^+_{R,0} V_{0,3/2} \ket{0} &=& V_{1,0} \ket{0}, \nonumber\\
G^-_{L,0}G^+_{R,0}V_{0,3/2} \ket{0} &=& G^+_{R,0}G^-_{L,0} V_{0,3/2}\ket{0} = V_{0,-3/2}\ket{0}. \nonumber
\end{eqnarray}

\begin{figure}[h!]
\begin{center}
\resizebox{.8\textwidth}{!}{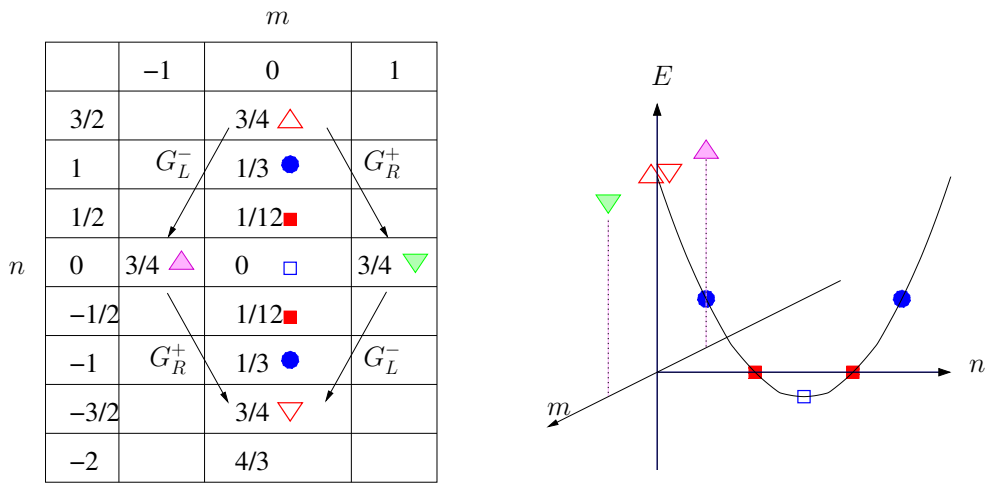}
\caption{The table shows $h_L + h_R = 3/4 m^2 + 1/3 n^2$, where $(-1)^{m+2n}$ is $+1$ in the NS and
$-1$ in the R sector. The big arrows point at the superpartners of the state with $h_L + h_R =3/4$. The symbols can
be found in the 3D plot, where the energy $E=h_L + h_R-c/12$ is plotted against $n$ and $m$. \label{fig:SCFT}}
\end{center}
\end{figure}

A pictorial summary of the above can be found in figure \ref{fig:SCFT}.

\section{Relation to the lattice model}\label{sec:lattice}

\subsection{Relation between sectors and boundary conditions}\label{sec:length3jpm1}

As was noted before, the supersymmetric structure as we discussed it in section \ref{sec:model} is only fully realized in the Ramond sector of the superconformal algebra. It turns out, however, that the NS is also realized in the lattice model, namely when we impose anti-periodic boundary conditions. In particular, we find from numerics that the lowest energy state for the supersymmetric model on a periodic chain with anti-periodic boundary conditions and length $L=0 \mod 3$, has a negative energy. A scaling analysis shows (see section \ref{sec:finitesizescaling}) that the energy of this state is $E=-1/12$, corresponding to the NS vacuum $\ket{0}$.

In the previous sections we have identified the Ramond sector of
the field theory to correspond to the lattice model on a chain
with length $L=0\mod 3$ with periodic boundary conditions. For the
chain of length $L=3j$, the fermion number in the ground state
sector is $f_{GS}=L/3=j$. For a chain of length $L=3j\pm1$, we
would correspondingly find $f_{GS}=L/3=j\pm1/3$. We know, however,
that for a chain of length $L=3j\pm1$ the ground state has fermion
number $f_{GS}=j$. It follows that, compared to the chain of
length $L=3j$, the chain of length $L=3j\pm1$ has a slightly
lower/higher charge density in the ground state sector. Now
remember that $m$ gives the charge compared to the charge in the
ground state sector of the chain of length $L=3j$, that is $L/3$.
Suppose that instead of $m=f-f_{GS}$, we now write $m=f-L/3$. For
the chain of length $L=3j$, the two definition are completely
equivalent. However, for a chain of length $L=3j\pm1$ we now find
that $m=f_{GS}-L/3=\mp1/3$ in the ground state sector. The
supersymmetry in the lattice model tells us that the chain with
periodic boundary conditions corresponds to the Ramond sector in
the field theory. Now that $m$ takes values in $\mathbb{Z}\mp1/3$,
it follows that in the Ramond sector, which has $3m/2+n \in
\mathbb{Z}+1/2$, $n$ is now integer. In fact, we find that adding
or subtracting one site from a chain of length $L=3j$ corresponds
in the field theory to acting with the operator $V_{\mp 1/3,-
1/2}$. Upon acting with this operator the highest weight states in
the Ramond sector become
\begin{eqnarray}
 V_{\mp 1/3,-1}\ket{0},  V_{\mp 1/3,0}\ket{0} \textrm{ and } V_{\mp 1/3,1}\ket{0},
\end{eqnarray}
for the chains of length $L=3j\pm1$. The corresponding energies
are respectively $E=1/3,0$ and $1/3$. This agrees with the fact
that these chain lengths only have one zero energy ground state.

\subsection{Lattice operators: fermion number and momentum}\label{sec:latticeoperators}

From section \ref{sec:spectrum} we know that the operator $F=J_{L,0}-J_{R,0}$ satisfies the same commutation relations with the supercharges as the fermion number operator in the lattice model. Using the definition of the $U(1)$ charges (\ref{eq:U1charges}), we find that this operator gives $q_L-q_R=m$ for a state $V_{m,n}\ket{0}$. If we identify the states $V_{0,\pm1/2}\ket{0}$ with the two zero energy ground states of the supersymmetric model on the one dimensional periodic chain with length $L=0 \mod 3$, which have fermion number $f_{GS}=L/3$, we can identify the fermion number operator with
\begin{eqnarray}\label{eq:fermlatscft}
 F=J_{L,0}-J_{R,0}+f_{GS}.
\end{eqnarray}

In the field theory the operator that generates translations in the space direction corresponds to rotations on the complex plane. The momentum operator on the lattice is thus proportional to $L_{L,0} - L_{R,0}$. All highest weight states in the field theory have $h_L=h_R$, which would imply zero momentum. However, the two ground states of the periodic chain of length $L=3j$ have momenta $p_0=\pm \pi/3 + \pi f_{GS} \mod 2 \pi$ \cite{fendley-2003-36}.

Using the fact that a boundary twist in the lattice model corresponds to a spectral flow (see also section \ref{sec:spectralflow2}) in the field theory, we can identify the momentum of a highest weight state with the sum of its $U(1)$ charges.

A boundary twist in the lattice model is defined by the condition that the wavefunction picks up a factor $\exp(2\pi \imath \alpha)$ when a fermion hops over the end of the chain, i.e. between site $L$ and site $1$. For $\alpha=0$ we have periodic boundary conditions, which corresponds to the Ramond sector, whereas for $\alpha=1/2$ we have anti-periodic boundary conditions, corresponding to the Neveu-Schwarz sector. Consequently, the boundary twist in the lattice model corresponds to a spectral flow in the continuum theory.

In the following we will show, on the one hand, that the momentum of a state in the lattice model depends linearly on the boundary twist and, on the other hand, that the index $n$ of a vertex operator $V_{m,n}$ will change linearly under spectral flow. These observations will allow us to relate the two.

Momentum, $p \mod 2 \pi$, can be defined by writing the eigenvalues of the translation operator as $t=e^{\imath p}$.
The boundary twist can be implemented by replacing the term that hops a particle over the boundary $c^{\dag}_L c_1+$ h.c. by $e^{2 \pi \imath \alpha}(c^{\dag}_L c_1+$ h.c.$)$. The eigenvalues of the translation operator for general $\alpha$ then follow from
\begin{eqnarray}
T_{\alpha}^L \ket{\psi} &=& e^{\imath p_0 L} e^{2 \pi \imath \alpha f} \ket{\psi} \equiv e^{ \imath p_{\alpha} L}\ket{\psi},
\end{eqnarray}
where $p_0$ is the momentum of $\ket{\psi}$ for $\alpha=0$, $L$ is the length of the system and $f$ is the total number of particles in the state $\ket{\psi}$. It follows that momentum indeed depends linearly on the boundary twist: $p_{\alpha}= p_0 + 2 \pi \alpha f/L \mod 2 \pi$.

It the continuum theory, the spectral flow is induced by the operator $V_{0,\alpha}$. Note that it conserves the fermion number, but changes the sector of the theory: $(-1)^{m+2n} \rightarrow (-1)^{m+2n} (-1)^{2\alpha}$.

If we now combine the fact that $p_{\alpha}= p_0 + 2 \pi \alpha f_{GS}/L + 2 \pi \alpha m/L \mod 2 \pi$ for a state in the lattice model and $n=n_0+\alpha$ for an operator in the field theory, we find that $p$ is proportional to $n$. By eliminating $\alpha$ and using the known results for the zero energy ground states in the lattice model we obtain
\begin{eqnarray}
 p=2 \pi n/3 + 2 \pi n m /L + f_{GS} \pi \mod 2 \pi.
\end{eqnarray}
Note that the middle term is precisely $2 \pi(h_R-h_L)/L$.

For the Ramond vacua $V_{0,\pm 1/2} \ket{0}$ we can easily check this relation. We know that the two ground states of the periodic chain of length $L=3j$ have momenta $p_0=\pm \pi/3 + \pi f_{GS} \mod 2 \pi$. Since the Ramond vacua have $n=\pm1/2$ and $m=0$, we find that this indeed nicely agrees with the equation above. Furthermore, one readily verifies that the ground state of chains of length $L=3j \pm 1$, which we identified in the previous section with the field $V_{\mp 1/3, 0}$, also has the correct momentum; $p=0+j\pi \mod 2 \pi$.

Finally, using the definition of the $U(1)$ charges (\ref{eq:U1charges}) we can write momentum as
\begin{eqnarray}\label{eq:momvallatscft}
 p = (q_L+q_R) \pi + 2 \pi(h_R-h_L)/L + f_{GS} \pi \mod 2 \pi
\end{eqnarray}
and we find that the momentum operator on the lattice can be expressed as
\begin{eqnarray}\label{eq:momlatscft}
 P&=& (J_{L,0}+J_{R,0})\pi + (L_{L,0} - L_{R,0}) 2\pi/L +f_{GS} \pi \mod 2 \pi . \nonumber\\
%&=& 2 J_{L,0} \pi + (L_{L,0} - L_{R,0}) 2\pi/L +f_{GS}\pi \mod 2 \pi.
\end{eqnarray}
% It may seem surprising that momentum depends on $J_{L,0}$ and not $J_{R,0}$, however, there is an ambiguity in our choice for the momentum interval. If we would choose $P \in [-\pi,\pi)$ instead of $[0,2 \pi)$, its expression would be symmetric in $J_{L,0}$ and $J_{R,0}$.

\subsection{Finite-size scaling: the Fermi velocity}
The finite-size scaling of the energy depends on the boundary conditions. For (anti-) periodic boundary conditions,
which corresponds to the cylinder on the field theory side, the scaling is given by \cite{Blote86,Affleck86}
\begin{eqnarray}
E_{\textrm{num}}=2 \pi E_{\textrm{SCFT}} v_F/N +\mathcal{O}(1/N^2),
\end{eqnarray}
where $N$ is the length of the finite system and $v_F$ the Fermi velocity. It follows that by comparing the finite size spectra with the spectrum of the field theory one can extract the Fermi velocity. In this case, however we can also obtain the Fermi velocity using the mapping of the supersymmetric model onto the XXZ chain. For the XXZ chain (\ref{eq:XXZ}) the Fermi velocity is given by
\begin{eqnarray}
v_F(\Delta)= \pi \sin \theta/\theta,
\end{eqnarray}
with $\cos \theta=-\Delta$. The supersymmetric model maps to the XXZ chain with $\Delta=-1/2$, so we find $\theta=\pi/3$ and the Fermi velocity of the corresponding XXZ chain is $v_F=(3 \sqrt 3)/2$. To find the Fermi velocity for the supersymmetric model, we note that the length $N$ of the XXZ chain is related to the length $L$ of the supersymmetric chain via $N=L-f$, where $f$ is the number of fermions in the supersymmetric model \cite{fendley-2003-36}. In the continuum limit the low energy states have approximately $f=L/3$, so $N=2L/3$. Combining all this, we find for the supersymmetric model that
\begin{eqnarray}
E_{\textrm{num}}=2 \pi E_{\textrm{SCFT}} v_F/L+\mathcal{O}(1/L^2),
\end{eqnarray}
with Fermi velocity $v_F=3/2 v_{F,XXZ}=(9 \sqrt{3})/4$.

%\emph{Luttinger parameters?}
\subsection{Finite size spectra}\label{sec:finitesizescaling}
In this section we analyze the numerically obtained spectra for
the supersymmetric model on the chain with periodic and
anti-periodic boundary conditions. We consider lengths up to
$L=27$.

\begin{figure}[htb]
  \centering
  \begin{minipage}[c]{0.48\textwidth}
    \centering
    \includegraphics[width=\textwidth]{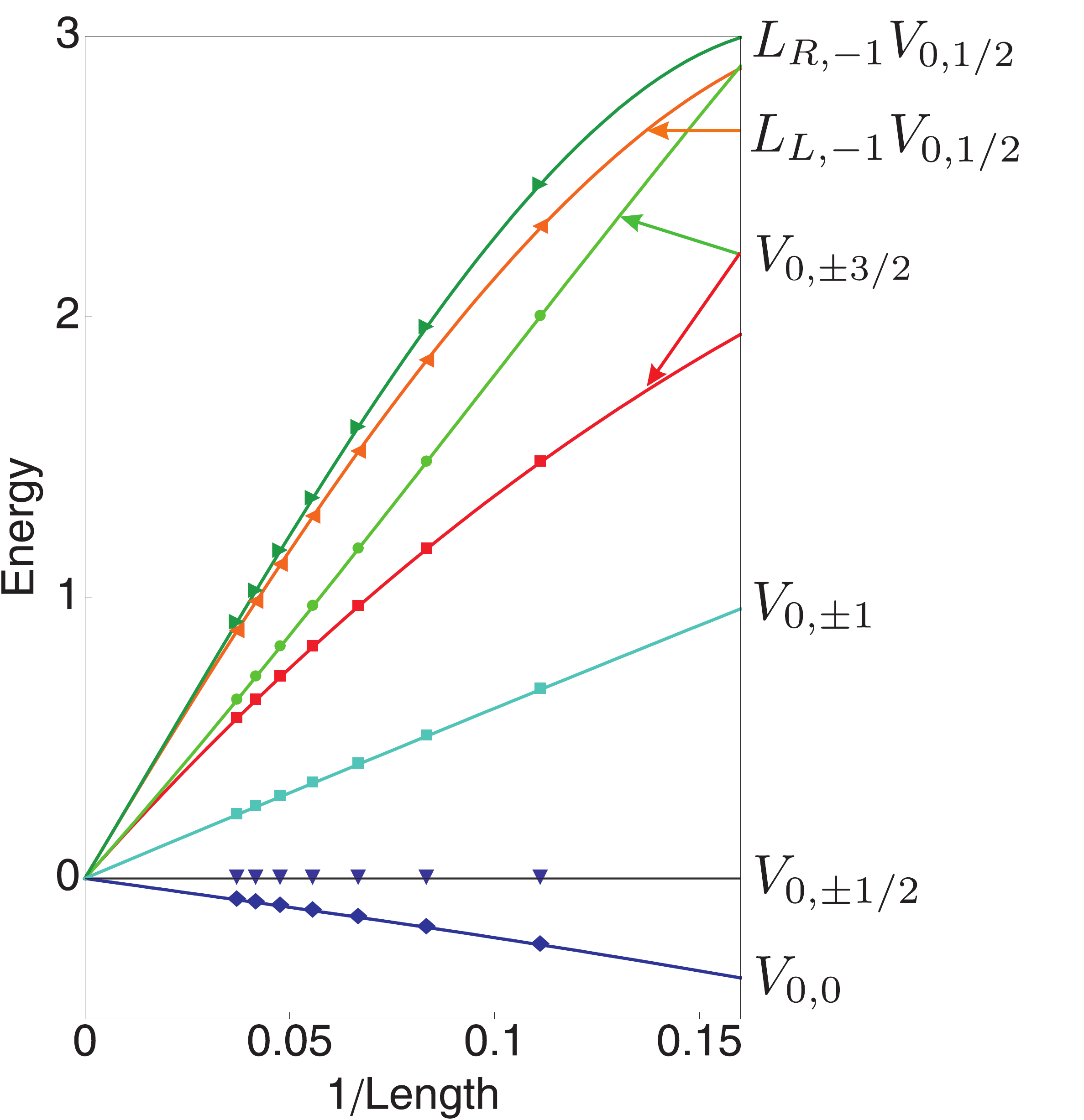}
  \end{minipage}
  \begin{minipage}[c]{0.48\textwidth}
\beq\nonumber
\begin{array}{|c|c|c|c|}
\hline
 \text{BC} & \text{field} & E_{\text{SCFT}} & \left.E_{\text{num}}\text{L/(2}\text{$\pi $v}_F\right) \\
\hline
 R & V_{0,-\frac{1}{2}} & 0 & 0 \\
 R & V_{0,\frac{1}{2}} & 0 & 0 \\
 R & V_{0,\pm \frac{3}{2}} & \frac{2}{3} & 0.626; 0.702 \\
 R & L_{L,-1} V_{0,\frac{1}{2}} & 1 & 1.001 \\
 R & L_{R,-1} V_{0,\frac{1}{2}} & 1 & 1.001 \\
 \text{NS} & V_{0,0} & -\frac{1}{12} & -0.083 \\
 \text{NS} & V_{0,1} & \frac{1}{4} & 0.25 \\
 \text{NS} & V_{0,-1} & \frac{1}{4} & 0.25\\
\hline
\end{array}
\eeq
  \end{minipage}
  \caption{On the left we plot the energy versus the inverse chain length for chains of length $L=3j$ with periodic (R) and anti-periodic (NS) boundary conditions. The numerical data and fits are shown. The fit function is $f(L)=a/L+b/L^2+c/L^3$. On the right, the table shows the results extracted from the scaling fits.}
     \label{fig:allscaling}
\end{figure}

For chains of length
$L=3j$ with anti-periodic bc we find that there is one negative energy state with fermion
number $f=L/3$. We identify this state with the vacuum of
the superconformal field theory. The
vacuum state has conformal dimensions $h_L=h_R=0$, so its energy is given by $E_{\textrm{SCFT}}=h_L+h_R-c/12=-1/12$, where we used $c=1$. In figure \ref{fig:allscaling} we show a scaling analysis of the numerically obtained energies of this
state for different lengths of the system. Remember that the scaling is given
by \cite{Blote86,Affleck86}
\begin{eqnarray}\label{eq:energy}
E_{\textrm{num}}=2 \pi E_{\textrm{SCFT}} v_F/L +\mathcal{O}(1/L^2),
\end{eqnarray}
where $v_F=9\sqrt{3}/4$. Using $E_{\textrm{SCFT}}=-1/12$, we find that the
energy of these states scales as $E_{\textrm{num}}=- 3 \sqrt{3}
\pi/(8L)\approx -2.041/L$. The function that gives the best fit to
the numerics is $f(L)=a/L+b/L^2+c/L^3$ with $a=-2.038$, $b=-0.056$ and $c=-6.509$.
Clearly, the value of $a$ agrees well with the theoretical value.

Figure \ref{fig:allscaling} also shows scaling analyses for other low lying levels of chains of length $L=3j$. Clearly, the results match well with the continuum theory. Similar results are obtained for chains of length $L=3j\pm1$ \cite{HuijseT10}.

\begin{figure}[h!bt]
\centering
\includegraphics[width=0.95\textwidth]{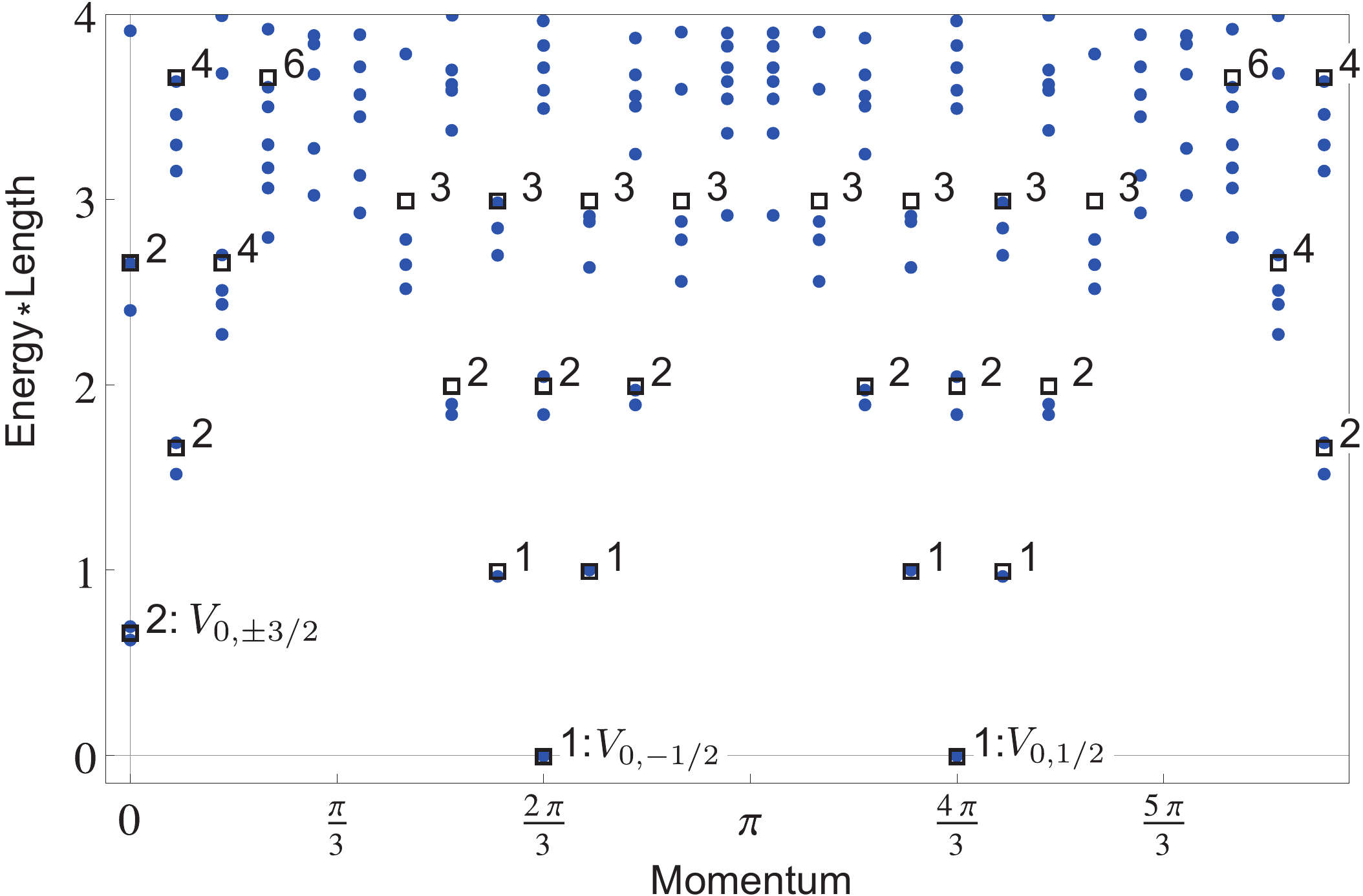}
\caption{We plot the energy versus momentum for the periodic chain of length $L=27$ at fermion number $f=9$ (filled dots). The numerically obtained energies are rescaled with the factor $2 \pi v_F/L$. The spectrum for the continuum theory is indicated by the open squares, where the number labels the degeneracy of the energy level. For the four lowest energy states the corresponding vertex operators are shown explicitly. In table \ref{tab:desc} one finds the operators corresponding to the descendants of the highest weight state $V_{0,1/2}\ket{0}$. \label{fig:plotN27f9alpha0v3}}
\end{figure}

\begin{table}
 \caption{In this table we show the descendants of the highest weight state $V_{0,1/2}\ket{0}$. This is the zero energy state in the Ramond sector with momentum $4 \pi/3$. We order the descendants by their energy, also called level, in the first column. The energy is given by $E=h_L+h_R-1/12$. In the second column we give $h_L-h_R$, which is directly related to the change in momentum with respect to the highest weight state. In the third column we give all the fields at the given energy and momentum. Finally, in the last column we give the superpartners of the fields at the given energy and momentum. Between brackets we indicate the relative charge, $m$, of the superpartners. The relative charge gives the change in fermion number with respect to the highest weight state. \label{tab:desc}}
\begin{center}
\begin{tabular}{|l|l|l|l|}
\hline
Energy & Momentum & Fields & Superpartners (charge) \\
\hline
0 & 0    & $V_{0,1/2}$ & - \\
\hline
1 &  1    & $L_{L,-1}V_{0,1/2}$ & $G^-_{L,-1}V_{0,1/2}= V_{-1,-1}$ (-1)  \\
  &  -1   & $L_{R,-1}V_{0,1/2}$ & $G^+_{R,-1}V_{0,1/2}= V_{1,-1}$ (+1) \\
\hline
2 &   2    & $(L_{L,-1})^2 V_{0,1/2}$ & $L_{L,-1}G^-_{L,-1}V_{0,1/2}= L_{L,-1}V_{-1,-1}$ (-1) \\ 
  &        & $L_{L,-2}V_{0,1/2}$ & $G^+_{L,-2}V_{0,1/2} = V_{1,2}$ (+1)  \\ \cline{2-4}
  & 0      & $L_{L,-1}L_{R,-1}V_{0,1/2}$ & $L_{R,-1} G^-_{L,-1}V_{0,1/2}= L_{R,-1}V_{-1,-1}$ (-1) \\ 
  &        & $ G^+_{R,-1} G^-_{L,-1}V_{0,1/2} = V_{0,-5/2}$ & $L_{L,-1} G^+_{R,-1}V_{0,1/2}= L_{L,-1}V_{1,-1}$ (+1) \\ \cline{2-4}
  &  -2   & $(L_{R,-1})^2 V_{0,1/2}$ & $G^-_{R,-2}V_{0,1/2} = V_{-1,2}$ (-1) \\
  &        & $L_{R,-2}V_{0,1/2}$ &  $L_{R,-1}G^+_{R,-1}V_{0,1/2}= L_{R,-1}V_{1,-1}$ (+1)  \\ 
 \hline
3 &  3  & $(L_{L,-1})^3 V_{0,1/2}$  &  $(L_{L,-1})^2 V_{-1,-1}$ (-1) \\ 
  &     & $L_{L,-1}L_{L,-2}V_{0,1/2}$  &  $L_{L,-2} V_{-1,-1}$ (-1)  \\
  &     & $L_{L,-3}V_{0,1/2}$  &  $L_{L,-1} V_{1,2}$ (+1)   \\ \cline{2-4}
  &  1  & $L_{R,-1}(L_{L,-1})^2 V_{0,1/2}$  & $(L_{L,-1})^2 V_{1,-1}$ (+1)   \\
  &     & $L_{R,-1}L_{L,-2}V_{0,1/2}$  &    $L_{L,-2} V_{1,-1}$ (+1)  \\
  &     & $L_{L,-1} V_{0,-5/2}$  &  $L_{R,-1} V_{1,2}$ (+1)   \\
  &     &    &  $L_{R,-1} L_{L,-1} V_{-1,-1}$ (-1) \\
  &     &    &  $G^+_{L,-2} V_{1,-1}$ (+2)   \\ \cline{2-4}
  & -1  & $L_{L,-1}(L_{R,-1})^2 V_{0,1/2}$  &  $(L_{R,-1})^2 V_{-1,-1}$ (-1)  \\
  &     & $L_{L,-1}L_{R,-2}V_{0,1/2}$  &  $L_{R,-2} V_{-1,-1}$ (-1)    \\
  &     & $L_{R,-1} V_{0,-5/2}$  &  $L_{L,-1} V_{-1,2}$ (-1)  \\ 
  &     &    &  $L_{R,-1} L_{L,-1} V_{1,-1}$ (+1) \\
  &     &    &  $G^-_{R,-2} V_{-1,-1}$ (-2)   \\ \cline{2-4}
  & -3  & $(L_{R,-1})^3 V_{0,1/2}$   &  $(L_{R,-1})^2 V_{1,-1}$ (+1)  \\
  &     & $L_{R,-1} L_{R,-2}V_{0,1/2}$  &  $L_{R,-2} V_{1,-1}$ (+1)  \\
  &     & $L_{R,-3}V_{0,1/2}$  &  $L_{R,-1} V_{-1,2}$ (-1)  \\   
 \hline
 \multicolumn{4}{|c|}{$\dots$} \\
 \hline
\end{tabular}
\end{center}

\end{table}

In figure \ref{fig:plotN27f9alpha0v3} we plot the spectrum of the periodic chain of length $L=27$ as a function of momentum to illustrate that the analysis in the previous section allows us to identify all the low lying states with fields in the field theory and their descendants. To get a nice fit with the numerics we have rescaled the energies of the states in the field theory using (\ref{eq:energy}). Furthermore, the momenta follow from (\ref{eq:momvallatscft}). In the plot we show the vertex operators corresponding to the four lowest lying states explicitly. The other states are descendants of these states, which are obtained by acting with the supersymmetry and Virasoro generators on these states. As an example, table \ref{tab:desc} shows the operators corresponding to the descendants of the highest weight state $V_{0,1/2}\ket{0}$. Similarly, we can identify the low energy states of chains with anti-periodic bc and lengths $L=3j\pm1$ with states in different sectors of the field theory \cite{HuijseT10}.  

An interesting point is that for the periodic chain of length $L=3j$, the two states that correspond to the fields $V_{0,\pm 3/2}$ are not degenerate at finite size. The same is true for the states at level 1 which correspond to the fields $L_{L/R,-1}V_{0,\pm 1/2}$. Since the model is exactly solvable, we know that this must be a finite size effect and should thus vanish in the continuum limit (see also \cite{Yu92}). For the two states that correspond to the fields $V_{0,\pm 3/2}$, we checked this explicitly by verifying that the energy difference between the two states goes to zero faster than one over the length of the system. Indeed, we find that the energy difference scales as $a/L^2+b/L^3$, with $a=52$ and $b=-86$.

Remarkably, we find that the states with the higher energy have
superpartners at $f=j+1$, whereas the states with lower energy
have superpartners at $f=j-1$. This is probably best explained via the mapping to the XXZ chain. The supersymmetric model on a chain of length $L$ at $1/3$ filling corresponds to the XXZ chain of length $N$ at zero magnetization. The two chain lengths are related via $N=L-f$, where $f$ is the number of fermions in the supersymmetric model. It follows that the supercharges which add or remove a fermion in the supersymmetric model translate into operators on the spin chain which change the length of the chain. Since the energy scales as one over the length, a state with more particles in the supersymmetric model, which has a shorter length in the XXZ chain, will thus have a higher energy. Conversely, a state with less particles will have a lower energy.

\section{Spectral flow}\label{sec:spectralflow2}
In this section we will discuss the effect of the spectral flow operator on the states. In the lattice model the spectral flow operator corresponds to the boundary twist operator. This correspondence
has proven to be very powerful in identifying critical modes in
supersymmetric models on ladders \cite{Huijse08b}. Since the supersymmetric model
on the chain is exactly solvable, we know that the spectral flow
should correctly describe the boundary twist. We have already seen
that the scaling behavior of the finite size spectra nicely
corresponds to the behavior one expects from the field theory.
However, for more complicated systems extracting the scaling
behavior can be very challenging, whereas boundary twists are easily
carried out. For this reason we will discuss the spectral flow and
boundary twist for the chain here in quite some detail.

The spectral flow is a map between different representations of the superconformal algebra. The different representations are characterized by the parameter $\alpha$ which appears in the mode expansion of the supercharges:
\begin{eqnarray}\label{eq:modeexpGJ}
 G^{\pm}(z) &=& \sum_r G^{\pm}_r z^{-r-3/2} \nonumber
\end{eqnarray}
where $r$ runs over all values in $\mathbb{Z}+\alpha$.  It can be shown
that the generators of the left- and rightmoving superconformal algebras transform
as follows under spectral flow \cite{Schwimmer87}
\begin{eqnarray}
 L^{\alpha}_{L/R,n}  &=& L^{0}_{L/R,n} + \alpha J^{0}_{L/R,n} + \frac{c}{6} \alpha^2 \delta_{n,0} \nonumber\\
 J^{\alpha}_{L/R,n}  &=& J^{0}_{L/R,n} + \frac{c}{3} \alpha \delta_{n,0} \nonumber\\
 G^{\alpha,+}_{L/R,r}  &=& G^+_{L/R,r-\alpha} \nonumber\\
 G^{\alpha,-}_{L/R,r} &=& G^-_{L/R,r+\alpha}. \nonumber
\end{eqnarray}
One can check that for $\alpha$ integer the algebra maps back to itself. Furthermore, for $r$ integer and $\alpha= \frac{1}{2}$ the spectral flow maps the Ramond sector onto the Neveu-Schwarz sector. 

Similarly, we have the following relations for the conformal dimensions and the $U(1)$ charges:
\begin{eqnarray}
 h^{\alpha}_{L,R}  &=& h^{0}_{L,R} + \alpha q^{0}_{L,R} + \frac{c}{6} \alpha^2 \nonumber\\
 q^{\alpha}_{L,R} &=& q^{0}_{L,R} + \frac{c}{3} \alpha .
\end{eqnarray}
It follows that energy changes parabolically with $\alpha$ under spectral flow
\begin{eqnarray}\label{eq:spflowE}
 E_{\alpha}&=& E_{0} + \alpha (q_L+q_R) + \frac{c}{3} \alpha^2.
\end{eqnarray}
If we define $Q=q_L-q_R$ and $\tilde{Q}=q_L+q_R$, which are related to charge and momentum respectively, we find that under spectral flow
\begin{eqnarray}\label{eq:spflowQ}
 Q_{\alpha}&=& Q_0 ,\nonumber\\
 \tilde{Q}_{\alpha}&=& \tilde{Q}_0 + \frac{2c}{3} \alpha ,
\end{eqnarray}
that is $Q$ is invariant and $\tilde{Q}$ changes linearly with $\alpha$ under spectral flow.

Remember that in the lattice model we have $p_{\alpha}= p_0 + 2 \pi \alpha f/L \mod 2 \pi$ where $L$ is the length of the system and $f$ is the total number of particles.

To compare the numerical values we obtain for the energy of finite size systems of length $L$ with a boundary twist we use $E_{\textrm{num}} (\alpha)=2 \pi E_{\alpha} v_F/L$, where $v_F$ is the Fermi velocity. Using the linear relation between momentum in the lattice model and the twist $\alpha$, we can express the energy as a parabolic function of momentum
\begin{eqnarray}
E_{\textrm{num}}(p_{\alpha})&=&2 \pi E_{0} v_F/L + \frac{(p_{\alpha}-p_{0})\tilde{Q}_{0} v_F}{f} + \frac{c(p_{\alpha}-p_{0})^2 v_F L}{6 \pi f^2} \nonumber
\end{eqnarray}
It follows that in a finite size system we should be able to fit the energies to the following curve
\begin{eqnarray}
E_{\textrm{num}}(p_{\alpha})&=& a + b p_{\alpha} + d p_{\alpha}^2,
\end{eqnarray}
where the fit parameters $b$ and $d$ will satisfy
\begin{eqnarray}\label{eq:spflowfits}
b &=& \frac{\tilde{Q}_{0}v_F}{f} - \frac{c p_{0} L v_F}{3 \pi f^2}, \nonumber\\
d &=& \frac{c L v_F}{6 \pi f^2}.
\end{eqnarray}
If we combine this with $E_{\textrm{num}} (p_{\alpha=0})=2 \pi E_{0} v_F/L$, we have three equations for four parameters in the continuum theory: the central charge $c$, the energy in the Ramond sector $E_{0}$, the sum of the $U(1)$ charges $\tilde{Q}$ and finally the Fermi velocity $v_F$. It follows that from the energy dependence on a boundary twist, one can extract $c$, $E_{0}$ and $\tilde{Q}$ as functions of the Fermi velocity.

\subsection{Spectral flow in finite size spectra}\label{sec:spectralflowchain}

\begin{figure}[h!bt]
\centering
\includegraphics[width=.9\textwidth]{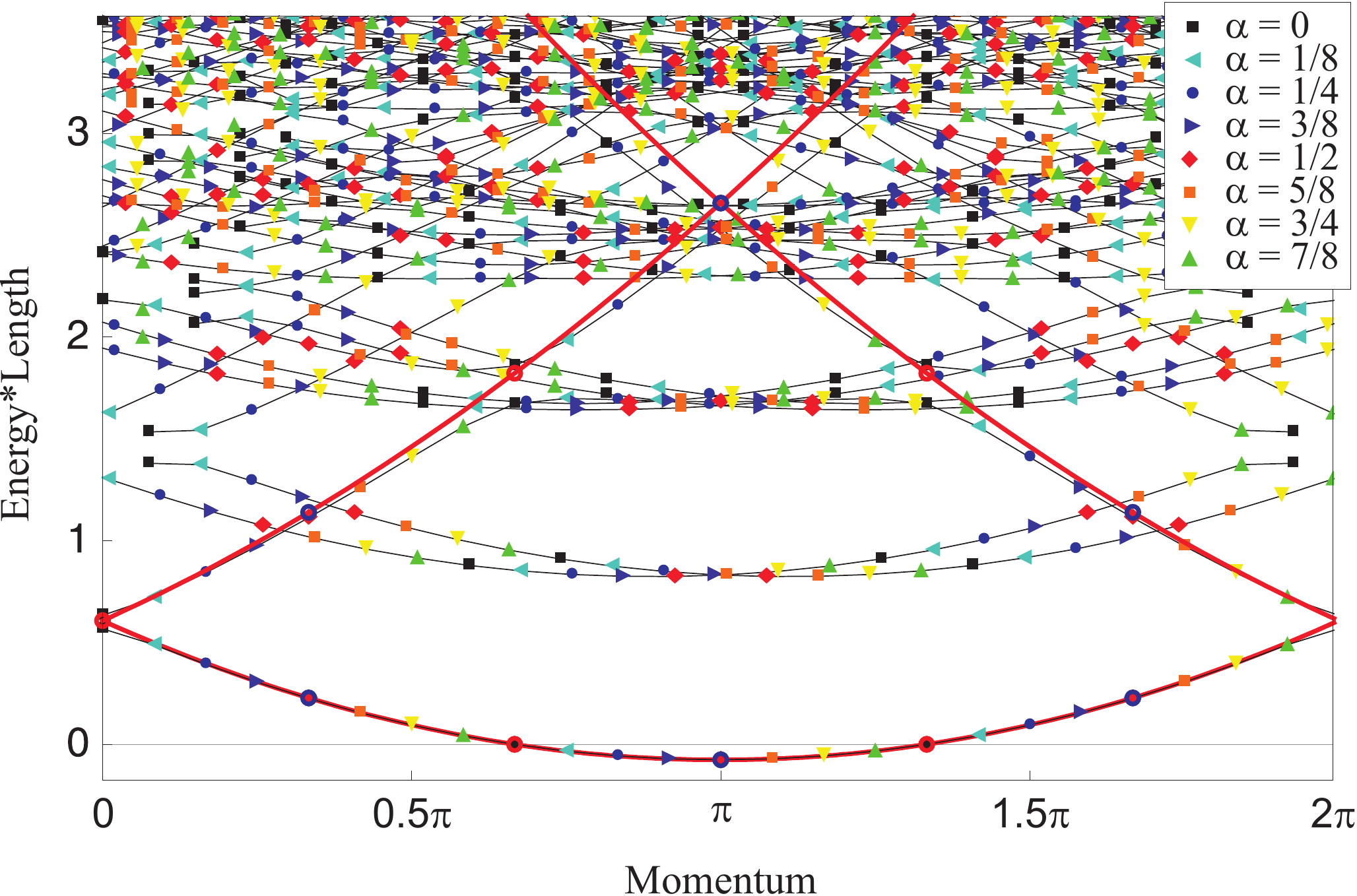}
\caption{We show the spectra (energy times chain length versus momentum) of the 27-site periodic chain with 9 fermions for $\alpha=0,\frac{1}{8},\frac{1}{4},\frac{3}{8},\frac{1}{2},\frac{5}{8},\frac{3}{4},\frac{7}{8}$. It follows that the spectrum given by the black squares is precisely the spectrum plotted in figure \ref{fig:plotN27f9alpha0v3}. The black line connects the levels for different values of the twist parameter. The red line is the parabola obtained from a fit to the flow of one of the Ramond vacua (the state corresponding to the field $V_{0,-1/2}$) to the Neveu-Schwarz vacuum (the state corresponding to the field $V_{0,0}$). The open red (blue) circles on this parabola correspond to $\alpha=0 \mod 1 $ ($\alpha=1/2 \mod 1 $). The two lowest energy states follow the parabola very nicely as a function of the twist parameter. For the higher energy states, we see that there are avoided level crossings (see also the end of section \ref{sec:finitesizescaling}) at integer values of $\alpha$, but for intermediate values of $\alpha$ they still qualitatively follow the parabola. By eye one can also clearly distinguish the parabola's through the first descendants of the Neveu-Schwarz vacuum, again interrupted by occasional avoided crossings. \label{fig:spectralflowL27}}
\end{figure}

In this section we present the data for periodic chains of lengths up to $L=27$ with a boundary twist.
We compute the spectrum of the system for $\alpha=\frac{s}{8}$, with $s=0,\dots,7$. For $s=0$ ($s=4$) we have periodic (anti-periodic) boundary conditions. An example is shown in figure \ref{fig:spectralflowL27}, for the chain of length $L=27$ and particle number $f=9$. For the low lying states one can easily see how the energy changes under the boundary twist. The drawn lines are parabolic fits to the energies $E_{\textrm{num}}$ as a function of the momenta $p_{\alpha}$. For the two lowest lying states, with $E_{\textrm{num}}=0$ for $p_{\alpha=0}=2\pi/3$ and $p_{\alpha=0}=4\pi/3$ we find the fits $f(x)=a+bx+dx^2$, with $(a,b,d)=(0.607, -0.435, 0.069)$ and $(a,b,d)=(0.579, -0.422, 0.068)$ respectively. Using equations (\ref{eq:spflowfits}) and the usual finite size scaling for the energy, we find
\begin{eqnarray}
 \frac{E_{\alpha}}{c}&=&\frac{E_{\textrm{num}}(p_{\alpha}) L^2}{d 12 \pi^2 f^2} \\
\frac{\tilde{Q}_{\alpha}}{c}&=&\frac{b+2d p_{\alpha} L}{d 6 \pi f}.
\end{eqnarray}
For the first fit we find $(E_0/c,\tilde{Q}_0/c)=(0,-0.334)$ and $(E_{1/2}/c,\tilde{Q}_{1/2}/c)=(-0.083,0.000)$ and for the second fit we find $(E_0/c,\tilde{Q}_0/c)=(0,0.334)$ and $(E_{1/2}/c,\tilde{Q}_{1/2}/c)=(0.254,0.675)$. It is clear that both fits correspond quite accurately with the theoretically predicted values of $(E_0/c,\tilde{Q}_0/c)=(0,\pm 1/3)$ in the Ramond sector and $(E_{1/2}/c,\tilde{Q}_{1/2}/c)=(-1/12,0)$ and $(E_{1/2}/c,\tilde{Q}_{1/2}/c)=(1/4,2/3)$ in the Neveu-Schwarz sector.

Note that the two fits are almost the same, as they should since the fields $V_{0,-1/2}$ and $V_{0,1/2}$ flow into each other under spectral flow, so their energies lie on the same parabola.

In table \ref{tab:spflowchain} in appendix \ref{app:spflow} we summarize the values we extract
from the parabola fits for $(E_{1/2}/c,\tilde{Q}_{1/2}/c)$ for
various system sizes. It is important to note, first of all, that
the values are quite accurate already for very small system sizes
and, second of all, that we do not have to compare systems of
different lengths. These two properties make this analysis very
powerful, also for systems for which we do not know what the
continuum limit is.

\section{Entanglement entropy}\label{sec:entangS}
Entanglement entropy is often used as a measure for the entanglement between two spatially separated parts of the system, but it is also a powerful technique to study criticality in finite size systems. Let $\rho$ be the density matrix of a system in a pure quantum state $\ket{\Psi}$: $\rho=\ket{\Psi}\bra{\Psi}$. Let us now divide the system in two parts $A$ and $B$, such that the Hilbert space can be written as $\mathcal{H}=\mathcal{H}_A \otimes \mathcal{H}_B$ and thus $\ket{\Psi}=\ket{\Psi_A}\ket{\Psi_B}$. We then define the reduced density matrix of subsystem $A$ as
\begin{eqnarray}
 \rho_A= \textrm{Tr}_B \rho.
\end{eqnarray}
The entanglement entropy is now defined as the Von Neumann entropy of the reduced density matrix
\begin{eqnarray}
 S_A = - \textrm{Tr} \rho_A \ln \rho_A,
\end{eqnarray}
and equivalently for $S_B$. For a system in a pure quantum state we have $S_A=S_B$.

For a system with a gap and thus a finite correlation length, the
entanglement entropy typically saturates at a certain value when
the size of the subsystem exceeds a certain length related to the
correlation length. For a critical system, which can be described
by a conformal field theory in the continuum limit, the
entanglement entropy does not saturate. In fact, it shows a
universal scaling law at a conformal critical point
\cite{Holzhey94,Vidal03,Calabrese04}
 \begin{eqnarray}
S(l_A)=\frac{c}{3} \ln (l_A) + b,
\end{eqnarray}
where $c$ is the central charge of the conformal field theory and $b$ is a non-universal constant. Finally, for a one dimensional quantum critical system of finite size $L$, the entanglement entropy scales as \cite{Calabrese04,Korepin04}
\begin{eqnarray}\label{eq:entangS}
S(l_A)=\frac{c}{3} \ln (\frac{L}{\pi}\sin(\frac{l_A \pi}{L})) + b,
\end{eqnarray}
which reduces to the expression above for $L \rightarrow \infty$.

It is now clear, that if one can compute the entanglement entropy of a one dimensional system, this can be a very powerful way of studying the system. It can be used to, first of all, determine whether the system is critical or not and, second of all, if it is critical, to determine the central charge of the continuum theory. An important boundary condition for this method to work, is that one is able to determine the entanglement entropy for a subsystem larger than the correlation length if the system is gapped. For exact diagonalization this is clearly not always the case. However, if the system is studied using density matrix renormalization group (DMRG) methods, this condition is often met. Moreover, the entanglement entropy comes out essentially for free if one determines the ground state of the system with DMRG.

For the supersymmetric model on the chain, the entanglement entropy has been studied using DMRG methods \cite{Campostrini}. The results were fitted very well by (\ref{eq:entangS}) and always in good agreement with central charge $c=1$ (with errors $\ll 1\%$) for the ground state of chains with periodic boundary conditions and $L=3j\pm1$ or with anti-periodic boundary conditions and $L=3j$. The ground state for the other cases, that is periodic boundary conditions and $L=3j$ or anti-periodic boundary conditions and $L=3j\pm1$, is degenerate. To determine the central charge reliably in these cases one would have to construct translational invariant ground states. Finally, for the chain with open boundary conditions it is difficult to obtain a precise determination of the central charge since the entanglement entropy is plagued by very strong oscillations of period 3. These oscillations are due to subleading finite size corrections to the entanglement entropy \cite{Laflorencie06,Cardy10}. Similar oscillations due to subleading corrections are observed in the fermion number densities (see section \ref{sec:obc2}).

\begin{figure}[h!]
     \centering
     \subfigure[$L=22$ and $f=7$.]
     {\includegraphics[width=.45\textwidth]{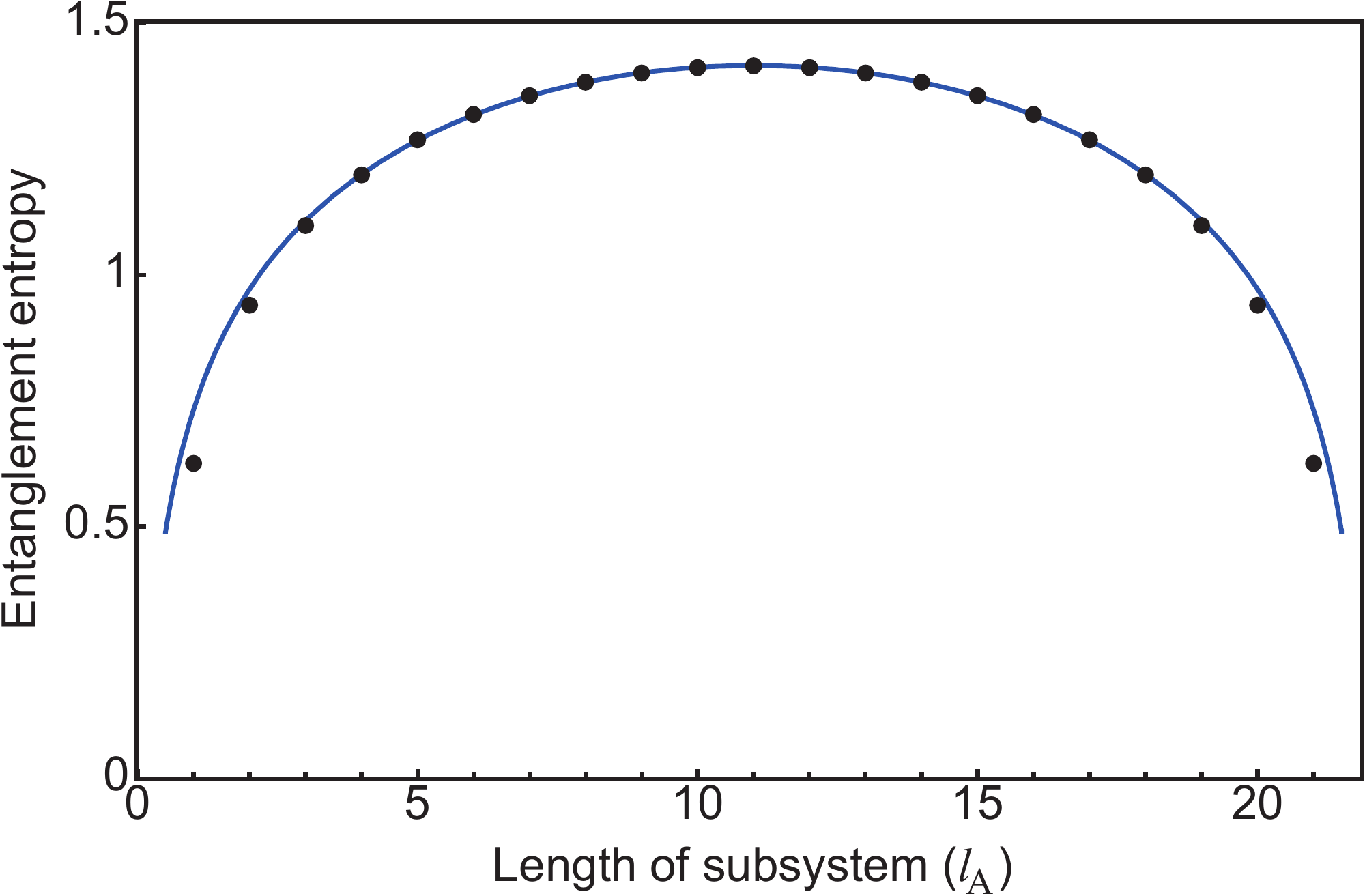}}
    \hspace{0.2cm}
     \subfigure[$L=23$ and $f=8$.]
     {\includegraphics[width=.45\textwidth]{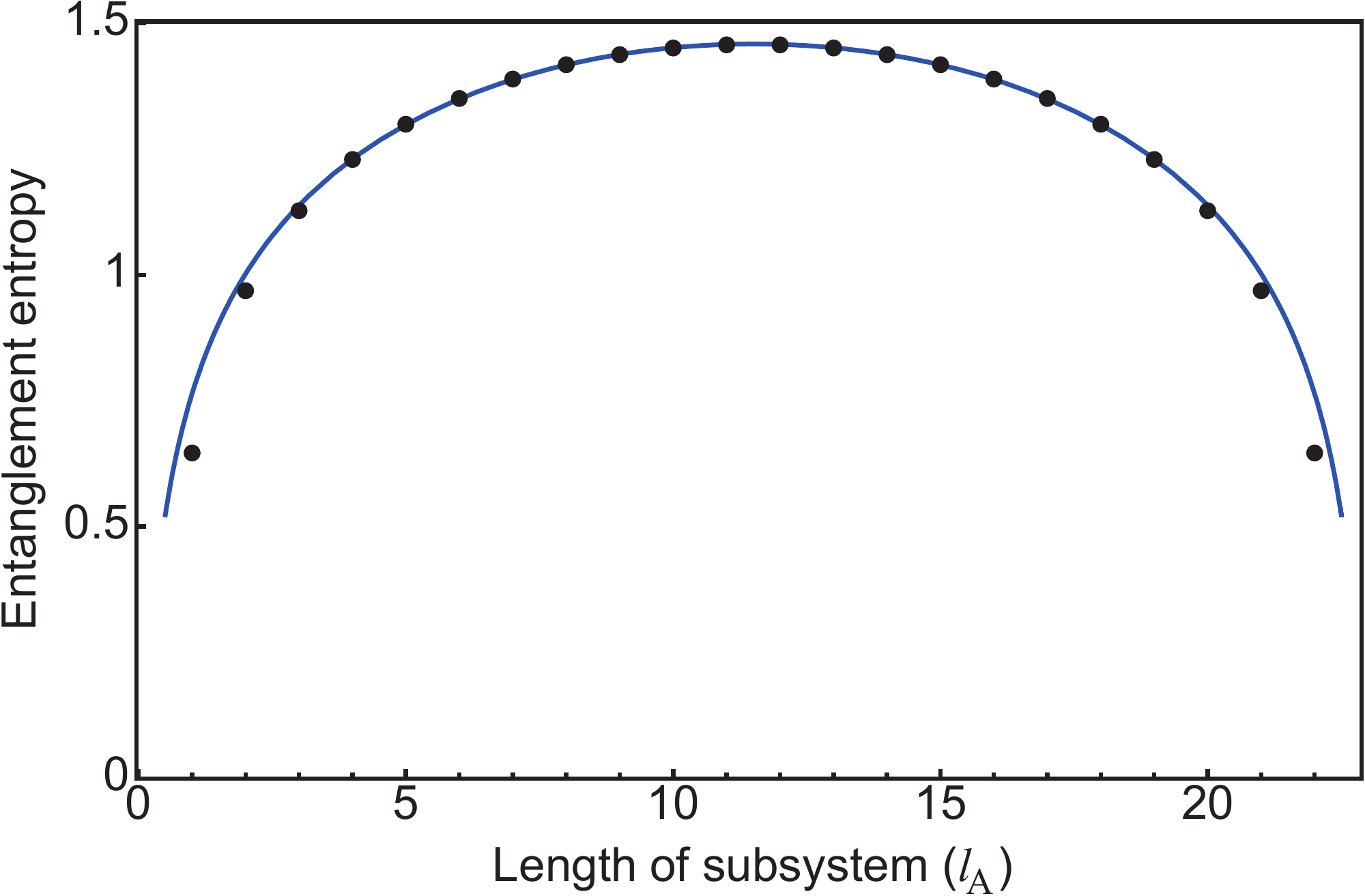}}
     \caption{Entanglement entropy versus subsystems size $l_A$ for the ground state of the chain of length $L=22$ with $f=7$ fermions (left) and the chain of length $L=23$ with $f=8$ fermions (right). \label{fig:entangSchain}}
\end{figure}

Using exact diagonalization, we determined the entanglement entropy for the ground state of chains with periodic boundary conditions and $L=3j\pm1$, for $L$ up to 23. The systems are too small to get a good determination of the central charge. They are, however, in reasonable agreement with $c=1$. Excluding the values for $l_A<3$ and $l_A>L-3$, we obtain $c\approx 1.05$ for $L=22$ and $c\approx 1.04$ for $L=23$ (see figure \ref{fig:entangSchain}).

Appart from a more thorough analysis of the entanglement entropy of this system, it would be interesting to look at the entanglement entropy of two disjoint intervals. It has recently been shown that such an analysis can  in principle reveal all scaling dimensions of the continuum theory \cite{Calabrese10}

\section{Open boundary conditions}\label{sec:openbc}
In this section we will discuss the chain with open boundary conditions. We identify the continuum theory (the $\mathcal{N}=2$ superconformal field theory at $c=1$) and the three sectors corresponding to the three possible chain lengths modulo three. 

\subsection{Continuum theory}\label{sec:obc1}
For open boundary conditions we do not expect the left- and right-moving modes to decouple, so the simplest guess for the continuum field theory is the $\mathcal{N}=2$ superconformal field theory with central charge $c=1$. By comparing the spectrum of this theory with numerical computations of finite size spectra, we concluded that this is indeed the correct guess.

The states of the field theory are given by the vertex operators
\begin{equation}\label{eq:vertexhol}
V_m=e^{(\imath m \Phi/\sqrt{3})},
\end{equation}
where the $\sqrt{3}$ comes from the compactification radius $R=\sqrt{3}$ and in the Ramond (NS) sector we have $m \in \mathbb{Z}+1/2$ ($m \in \mathbb{Z}$). The conformal
dimension $h_m$ corresponding to $V_m$ is
\begin{equation}
h_m=m^2/6.
\end{equation}
For $m=\pm3$ we find the supercharges, given by
\begin{equation}
G^{\pm}=e^{(\pm \imath \sqrt{3} \Phi)},
\end{equation}
with conformal dimension $h=3/2$.

In the following we only consider the Ramond sector, since this is
the sector that is realized in the lattice model. There are three
highest weight states that we need to consider. They correspond to
the primary fields: $V_{-1/2}$, $V_{1/2}$ and $V_{3/2}$. All other
states are generated by the supercharges and the Virasoro algebra.
The fields $V_{-1/2}$ and $V_{1/2}$ both have conformal dimension
$h=1/24$. Since the hamiltonian is given by $H=L_0-c/24$, they
correspond to zero energy states. The state $V_{3/2} \ket{0}$,
however, has energy $E=9/24-1/24=1/3$ and also has a superpartner
($V_{-3/2} \ket{0}$).

Since the supercharges change $m$ by $\pm 3$, we infer that $m/3$ corresponds to the fermion number in the lattice model. It thus quickly follows that the three sectors; $m=1/2,3/2,5/2 \mod 3$, are related to the three sectors in the lattice model with chain lengths $L=0,1,2 \mod 3$. Indeed we know (see e.g. \cite{Huijse08a}) that chains of length $L=0,2 \mod 3$ have one zero energy ground states and chains of length $L=1 \mod 3$ have all energies larger than zero. It follows that chains with length $L=1 \mod 3$ correspond to the sector with $m=3/2 \mod 3$. To identify the sector of the other two chain lengths, we look at the first excited state and its superpartner. The first excited state is given by $L_{-1} V_{\pm 1/2} \ket{0}$ and the respective superpartners are $G^{\mp}_{-1} V_{\pm 1/2} \ket{0}=V_{\mp 5/2} \ket{0}$. One easily checks that the superpartners indeed have energy $E=m^2/6-1/24=1$. The difference is that one occurs at $f=f_{GS}+1$ and the other at $f=f_{GS}-1$. If we compare this with the finite size spectra we quickly conclude that open chains with length $L=3j$ correspond to the sector with $m=5/2 \mod 3$ and chains with length $L=3j-1$ correspond to the sector with $m=1/2 \mod 3$.

Finally, we find the following relation between fermion number $f$, chain length $L$ and the quantum number $m$:
\begin{equation}
\tilde{f} \equiv f-L/3 = (m+1/2)/3.
\end{equation}
For $m=-1/2$ this relation gives $\tilde{f}=0$, which agrees with $f=j$ and $L=3j$. For $m=1/2$ this relation gives $\tilde{f}=1/3$, which agrees with $f=j$ and $L=3j-1$. Finally, for $L=3j+1$ the two lowest energy states are found at $f=j$ and $f=j+1$, which matches with $m=-3/2$ and $m=3/2$.

The spectrum per sector now simply follows from the highest weight states and their descendants. The character formula is given by
\begin{eqnarray}
 \chi_h (q) &=& q^{h} /\eta(q)\nonumber\\
&=& q^{m^2/6} /\eta(q),
\end{eqnarray}
where, as usual, $\eta(q) \equiv q^{c/24} \prod_{k=1}^{\infty} (1-q^k)$ and $q \equiv e^{2 \pi \imath \tau}$.
From this formula we obtain the degeneracies at each level. For the first few levels the degeneracies are summarized in table \ref{tab:deg}. The energy of the $n$-th level is given by $E=m^2/6-1/24+n$.

\begin{table}
\caption{The degeneracy at level $n$ is given by the number of partitions $p(n)$. The energy at level $n$ is given by $E=m^2/6-1/24+n$. \label{tab:deg}}
\begin{center}
\begin{tabular}{|l|l|l|l|l|l|l|l|l|l|l|l|l|l|}
\hline
level & 0 & 1& 2& 3& 4& 5& 6 & 7& 8& 9& 10& 11& 12 \\
\hline
degeneracy & 1& 1 & 2& 3& 5& 7& 11 & 15& 22& 30& 42& 56& 77 \\
\hline
\end{tabular}
\end{center}
\end{table}

\subsection{Finite size spectra}
\begin{figure}[h!tb]
     \centering
     \subfigure[The energy levels $E=m^2/6-1/24+n$ are plotted versus $m$ for $m=5/2 \mod 3$ which corresponds to chains of length $L=3j$. On the horizontal axis we also indicate the shifted fermion number $\tilde{f} \equiv f-L/3 = (m+1/2)/3$. The level corresponding to the highest weight state and the levels that are generated from this field by the supercharge operators, that is the levels with energy $E=m^2/6-1/24$, are indicated by a thick bar. The descendants are indicated by thinner bars. The labels indicate the degeneracy of the levels.\label{fig:openchain1}]
     {\resizebox{.45\textwidth}{!}{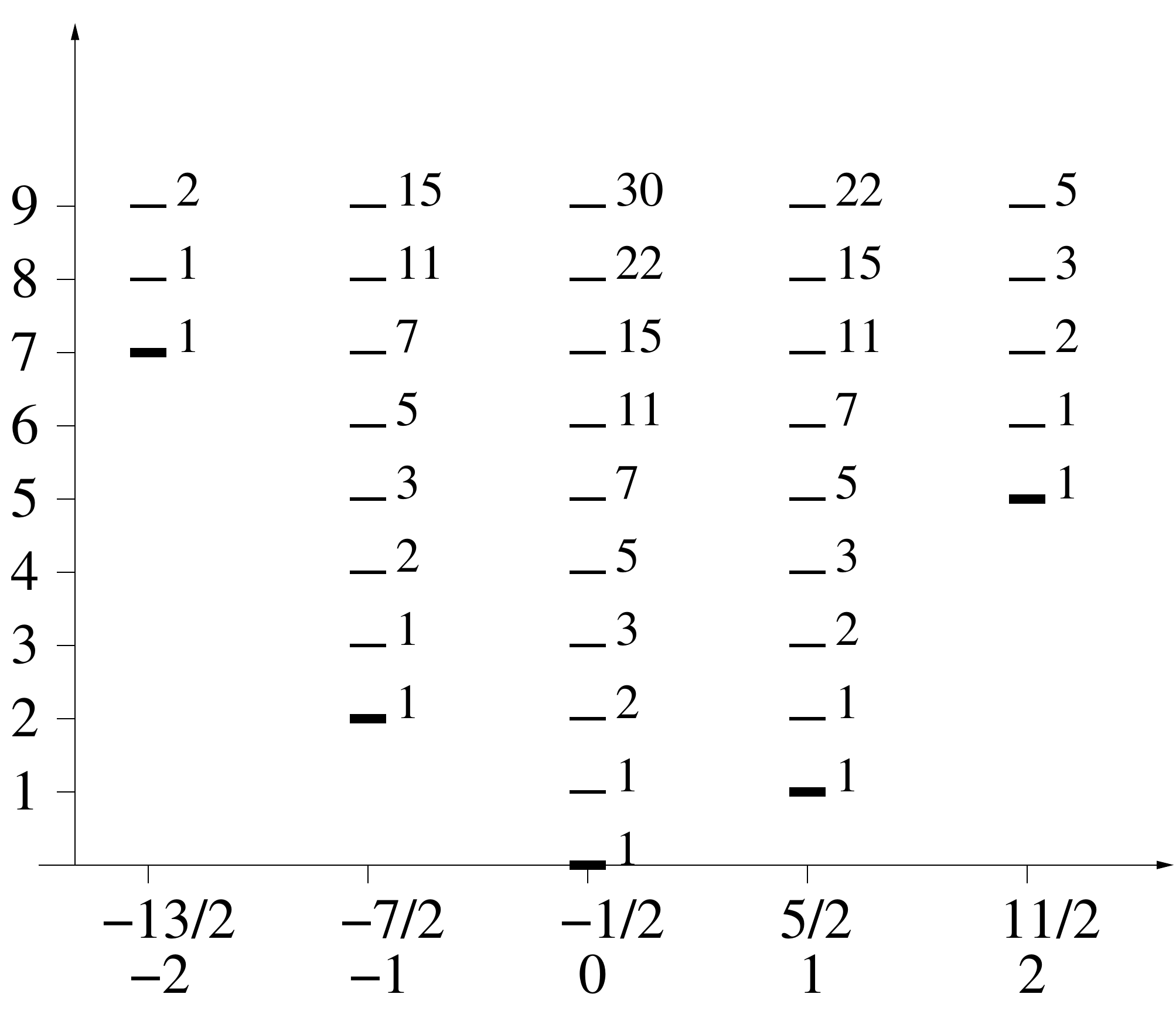}}
    \hspace{0.3cm}
     \subfigure[The numerically fitted values of the energy are plotted versus the shifted fermion number $\tilde{f} \equiv f-L/3$ for chains of length $L=3j$. The fit values are obtained by fitting the numerically obtained energies as a function of the chain length $L$ with the function $f(L)=a/L+b/L^2+c/L^3$. The energy is then given by $E=a/(\pi v_F)$. The labels indicate the number of overlapping data points. \label{fig:openchain1num}]
     {\includegraphics[width=.5\textwidth]{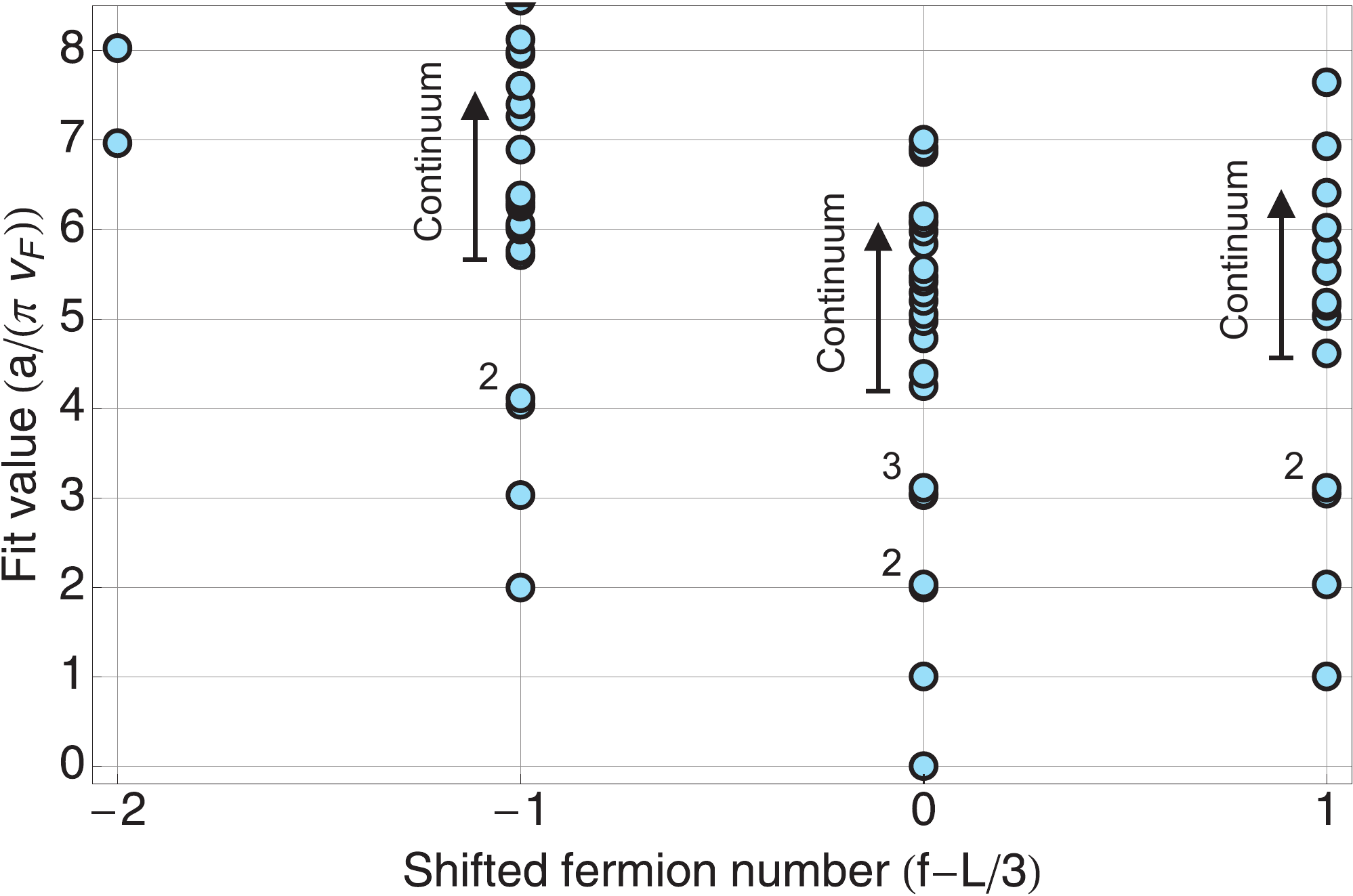}}
     \caption{On the left we show the theoretically predicted spectrum and on the right the spectrum obtained from fits to the numerically obtained spectra. These spectra are for chains of length $L=3j$.}
\end{figure}

We now compare the continuum spectrum to the numerics. We have performed a scaling analysis for the numerically obtained energies for chains of lengths up to $L=23$. The low lying levels are nicely fitted with the function $f(L)=a/L+b/L^2+c/L^3$ (for details see \cite{HuijseT10}). For open boundary conditions the scaling is given by
\begin{eqnarray}
 E_{\textrm{num}}=\pi E_{\textrm{SCFT}} v_F/L+\mathcal{O}(1/L^2),
\end{eqnarray}
where the Fermi velocity is given by $v_F=9\sqrt{3}/4$. It follows that the energies in the continuum limit can be extracted from the fits via $E=a/(\pi v_F)$. As an example we show the continuum limit spectrum that we extracted in this way for the chain of length $L=0 \mod 3$ in figure \ref{fig:openchain1num}, which should be compared to the continuum spectrum for $m=5/2 \mod 3$ plotted in figure \ref{fig:openchain1} (similar results are obtained for $L=1,2 \mod 3$  and can be found in \cite{HuijseT10}).

Clearly, for the first few levels, we find a very nice agreement with the theoretically obtained spectra. For the higher levels, the fits are not very reliable. We indicate two reasons for this. First of all, there is a large degeneracy of these levels in the continuum limit. However, in the finite size spectra the degeneracies are not realized, since for finite size matrices the eigenvalues tend to spread. Second of all, since there are corrections of order $1/L^p$ with $p\geq2$, there can be level crossings as a function of the length. Numerically, however, we simply connect the $n$-th level at different lengths, so we do not take possible level crossings into account. The latter argument also explains why supersymmetry may appear to be broken in the spectra extracted from the fits. Clearly, this is not the case in the original numerical spectra.

Using density renormalization group methods, one can obtain the low lying levels for much larger system sizes. For the 120-site chain with open boundary conditions it was found that \cite{Campostrini} the spectrum and level degeneracies are, at least up to level 7, in excellent agreement with the continuum theory in the sector with $m=5/2 \mod 3$.

\section{One-point functions}\label{sec:obc2}
In \cite{Beccaria05} Beccaria and De Angelis discuss the supersymmetric model on the chain with open boundary conditions and length $L \mod 3=0$. In particular, they use non standard number theoretical methods to obtain exact expressions for the ground state wave function, on the one hand, and on the other hand, they study the finite size scaling behavior of some simple correlation functions using exact diagonalization. Their results for the one point-function $\langle n_k\rangle =\bra{\psi_0} c^{\dag}_k c_k \ket{\psi_0}$ can be summarized as follows. They find that $\langle n_k\rangle $ has a clear $\mathbb{Z}_3$ substructure. The one-point functions $\langle n_{k,k=1\mod3}\rangle $ and $\langle n_{k,k=0\mod3}\rangle $ are not symmetric under $k \mapsto L-k$. The one-point function $\langle n_{k,k=2\mod3}\rangle $ \emph{is} symmetric under this map and shows a very different behavior from the other two.
% Second of all, they define $F_i = \sum_{k,k \mod 3=i} <n_k>$ as the average fermion number in the three branched distinguished by the site number modulo three. For these average fermion numbers they find
% \begin{itemize}
%  \item $F_0=F_1$ and $F_0+F_1+F_2=F$ with $F=L/3$,
%  \item $F_2=\frac{1}{2}F+\mathcal{O}(F^0)$ and $F_0=F_1=\frac{1}{4}F+\mathcal{O}(F^0)$ for large $L$.
% \end{itemize}
For the different branches they extract the following finite size scaling behavior
\begin{itemize}
 \item $\langle n_k\rangle -1/3 = f_+ \left( (k-k_+)/\tilde{L} \right) \tilde{L}^{-\nu}$ for $k \mod 3 =2$,
 \item $\langle n_k + n_{k+2}\rangle -2/3 = f_- \left( (k-k_-)/\tilde{L} \right) \tilde{L}^{-\nu}$ for $k \mod 3 =1$,
\end{itemize}
where $\tilde{L}=L/3+1$, $k_{\pm}=(L\pm1)/2$. They obtain the best fit for $\nu=0.33(2)$.

In the following we will use the observed $\mathbb{Z}_3$ substructure to propose an identification of the one-point functions with expectation values of operators in the superconformal field theory. 

Let us first identify the $\mathbb{Z}_3$ operator in the
superconformal field theory on the cylinder. Remember that the boson
is compactified\footnote{The normalisation in the Vertex operators is chosen such that the compactification radius $R=\sqrt{3}$ drops out.}: $\Phi \equiv \Phi + 2 \pi$ and $\tilde{\Phi} \equiv \tilde{\Phi} + 2 \pi$.
It follows that the operator $T$ that acts as follows $T:\Phi \mapsto \Phi + 2 \pi /3$ and $T:\tilde{\Phi} \mapsto \tilde{\Phi} + 2 \pi /3$ satisfies $T^3=1$. We now
consider the action of this operator on the vertex operators
$V_{0,\pm 1} = e^{ \pm \imath \tilde{\Phi}}$
\begin{eqnarray}
 T V_{0,\pm 1}= e^{ \pm \imath (\tilde{\Phi} + 2 \pi /3)} = e^{ \pm  2 \pi \imath /3} V_{0,\pm 1}.
\end{eqnarray}
Furthermore, we trivially have $T V_{0,0} = V_{0,0}$ and clearly $V_{0,0}$ is just the identity. It follows that $V_{0,0}$ and $V_{0,\pm 1}$ are eigenfunctions of the $\mathbb{Z}_3$ operator with eigenvalues $\omega_0=1$ and $\omega_{\pm 1}=e^{\pm 2 \pi \imath /3}$ respectively.

In the lattice the one-point function $\langle n_k\rangle $ shows strong oscillations with period three in $k$. It follows that we can identify the inverse translation operator which sends $k$ to $k+1$ as the $\mathbb{Z}_3$ operator\footnote{Note that there is an ambiguity here: we could just as well have identified the translation operator, which sends $k$ to $k-1$ with the $\mathbb{Z}_3$ operator. However, this ambiguity is fixed by comparison with the numerical results of \cite{Beccaria05}.}. We thus find that the functions,
\begin{eqnarray}
\omega_l n_{k-1} + n_k + \omega_l^{-1} n_{k+1},
\end{eqnarray}
are eigenfunctions of the inverse translation operator with eigenvalues $\omega_l= e^{2 \pi \imath l /3}$. In the following, we will assume\footnote{There is again an ambiguity here, since we could also take $k \mod 3=0,1$. This choice, however, is again fixed by comparison with the numerical results of \cite{Beccaria05} (see also the end of this section).} $k \mod 3=2$.

Combining these observations we find
\begin{eqnarray}
 A_0 V_{0,0} &=& n_{k-1} + n_k + n_{k+1} \nonumber\\
 A_{\pm1} V_{0,\pm 1} &=& e^{\pm 2 \pi \imath /3} n_{k-1} + n_k + e^{\mp 2 \pi \imath /3} n_{k+1}, \nonumber
\end{eqnarray}
where $A_0$ and $A_{\pm1}$ are constants. From the fact that the ground state has filling 1/3, we immediately find that $A_0=1$. Solving the above equations for the $n_k$ we obtain
\begin{eqnarray}
 3 n_k = V_{0,0} + A_1 V_{0,1} + A_{-1} V_{0,-1} \nonumber\\
 3 n_{k-1} = V_{0,0} + e^{-2\pi \imath/3} A_1 V_{0,1} + e^{2\pi \imath/3} A_{-1} V_{0,-1}\nonumber\\
 3 n_{k+1} =  V_{0,0} + e^{2\pi \imath/3} A_1 V_{0,1} + e^{-2\pi \imath/3} A_{-1} V_{0,-1} .\nonumber
\end{eqnarray}

In the following, we compute the expectation values of these operators on the cylinder and on the strip, which corresponds to closed and open boundary conditions respectively. For closed boundary conditions we compare our findings to analytical results presented in \cite{Fendley10b}. For open boundary conditions, we show how the expectation values can be computed analytically by mapping the strip onto the plane and introducing the mirror images to ensure the boundary conditions are preserved. The formulae we obtain are in nice agreement with the finite size scaling behavior found in \cite{Beccaria05}. An obvious follow-up on this work, is to extend it to two-point functions (see also \cite{Beccaria05}) and chains of length $L\neq 0 \mod 3$.

\subsection{Open boundary conditions}
The expectation values can be computed on the plane, where the correlator of vertex operators reads
\begin{eqnarray}\label{eq:correlator}
 \langle V_{m_1,n_1} (z_1,\z_1) \dots V_{m_k,n_k} (z_k,\z_k) \rangle = \prod_{i<j} (z_i-z_j)^{3\alpha_{L,i} \alpha_{L,j}/4}(\z_i-\z_j)^{3\alpha_{R,i} \alpha_{R,j}/4} .
\end{eqnarray}
Remember that for the vertex operators $V_{m,n}$ we have $\alpha_{L,R}=m\pm 2n/3$.

To illustrate how we go from the strip to the plane we compute the expectation value of the vertex operator $V_{0,1}$ in the Neveu-Schwarz vacuum. To compute this we have to do two steps. First we map the strip to the upper half plane using a conformal mapping, $z = e^{\pi \imath w/L}$, where $z$ and $w$ correspond to coordinates on the plane and the strip, respectively. Then we introduce a mirror image of the system on the upper half plane in the lower half plane, thus defining the system on the full complex plane, where we can compute the correlator. The first step gives
\begin{eqnarray}
 \bra{0} V_{0,1} (w,\w) \ket{0}_{\textrm{strip}} &=& \left( \frac{\partial z}{ \partial w}  \right)^{h_L}\left( \frac{\partial \z}{ \partial \w}  \right)^{h_R} \bra{0}  V_{0,1} (z,\z) \ket{0}_{\textrm{UHP}} \nonumber\\
 &=& \left( \frac{\pi \imath}{L} z  \right)^{1/6} \left( \frac{-\pi \imath}{L} \z  \right)^{1/6} \bra{0}  V_{0,1} (z,\z) \ket{0}_{\textrm{UHP}} ,\nonumber
\end{eqnarray}
where we used the fact that the conformal dimensions of $V_{0,1}$ are $h_L=h_R=1/6$. A correlator in the upper half plane can be computed using the image-technique. Choosing the boundary condition $\partial \Phi|_{\textrm{boundary}} = 0$, that is, there is no current flow across the boundary, the image-technique for a vertex operator gives
\beq
\bra{0}  V_{m,n} (z,\z) \ket{0}_{\textrm{UHP}}  = \bra{0}  V_{3\alpha_L/2} (z) V_{3\alpha_R/2} (z^*) \ket{0}, \nonumber
\eeq
and, in particular,
\beq
\bra{0}  V_{0,n} (z,\z) \ket{0}_{\textrm{UHP}}  = \bra{0}  V_{n} (z) V_{-n} (z^*) \ket{0}, \nonumber
\eeq
where the vertex operator with one index is purely holomorphic (see (\ref{eq:vertexhol})). In the end we set $z^*=\z$. We thus find
\beq
\bra{0}  V_{0,1} (z,\z) \ket{0}_{\textrm{UHP}} &=& \bra{0}  V_{1} (z) V_{-1} (z^*) \ket{0}\nn
&=& (z-z^*)^{-1/3}.
\eeq
Combining both steps and using $w=x+ \imath t$, we find
\beq
 \bra{0} V_{0,1} (x,t) \ket{0}_{\textrm{strip}} &=&  \left( \frac{\pi \imath}{L}  \right)^{1/3}  (- z \z)^{1/6} (z-\z)^{-1/3}\nn
&=& \left( \frac{\pi \imath}{L}  \right)^{1/3}  (-1)^{1/6} \frac{e^{\pi \imath (w-\w)/6L}}{(e^{\pi \imath w/L} - e^{-\pi \imath \w/L})^{1/3} }\nn
&=& (-1)^{1/6}  \left( \frac{\pi }{2 L}  \right)^{1/3} \sin^{-1/3}(\pi x/L) .\nonumber
\eeq 

\begin{figure}[h!]
     \centering
     \subfigure[One-point functions for chain length $L_c=30$. \label{fig:beccaria30}]
     {\includegraphics[height=5.2cm]{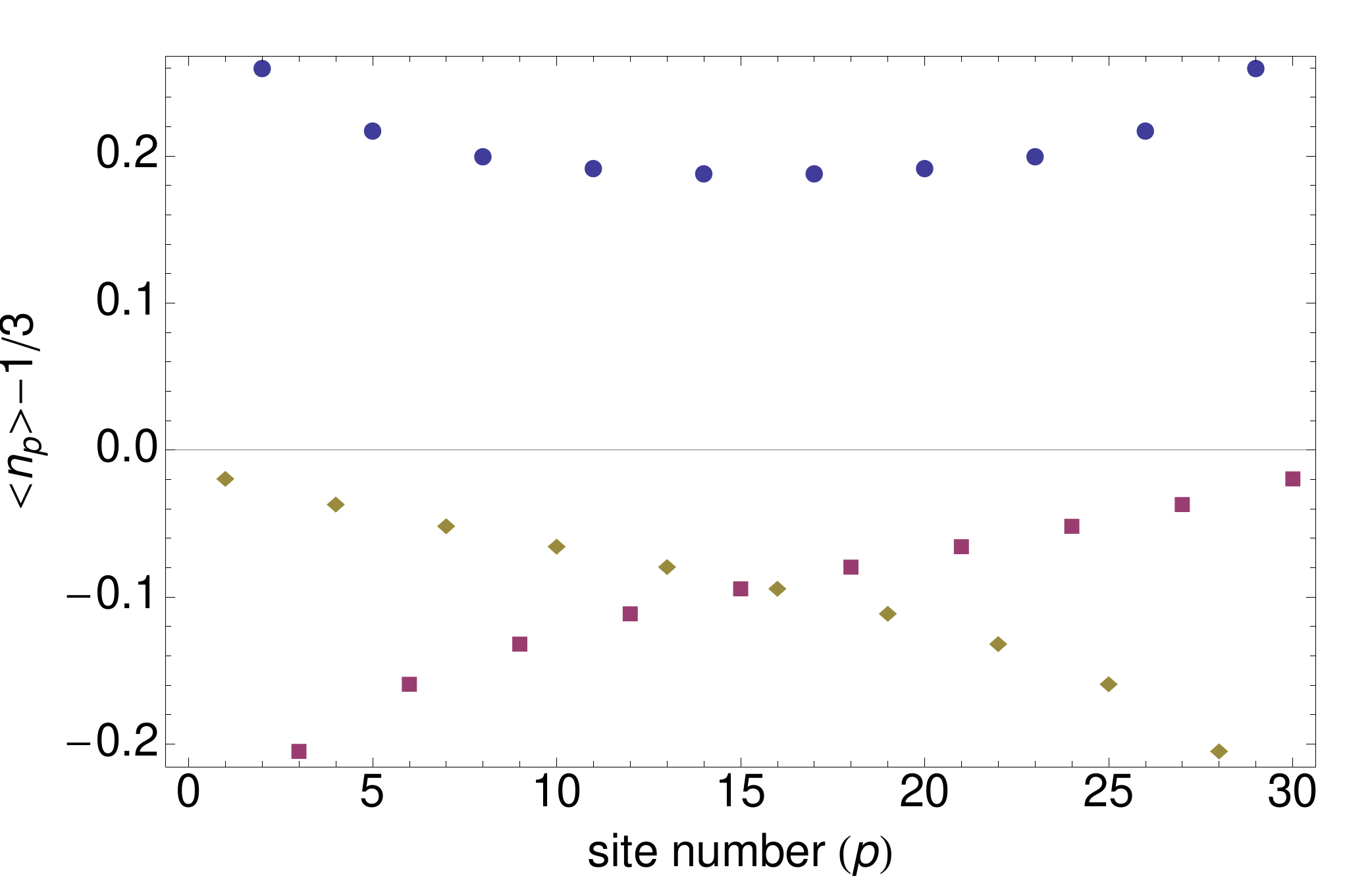}}
    \hspace{0.2cm}
     \subfigure[Finite size scaling functions versus the normalised position on the strip (which is related to the site number, $p$, via $x=p+1$). \label{fig:beccariafss}]
     {{\includegraphics[height=5.2cm]{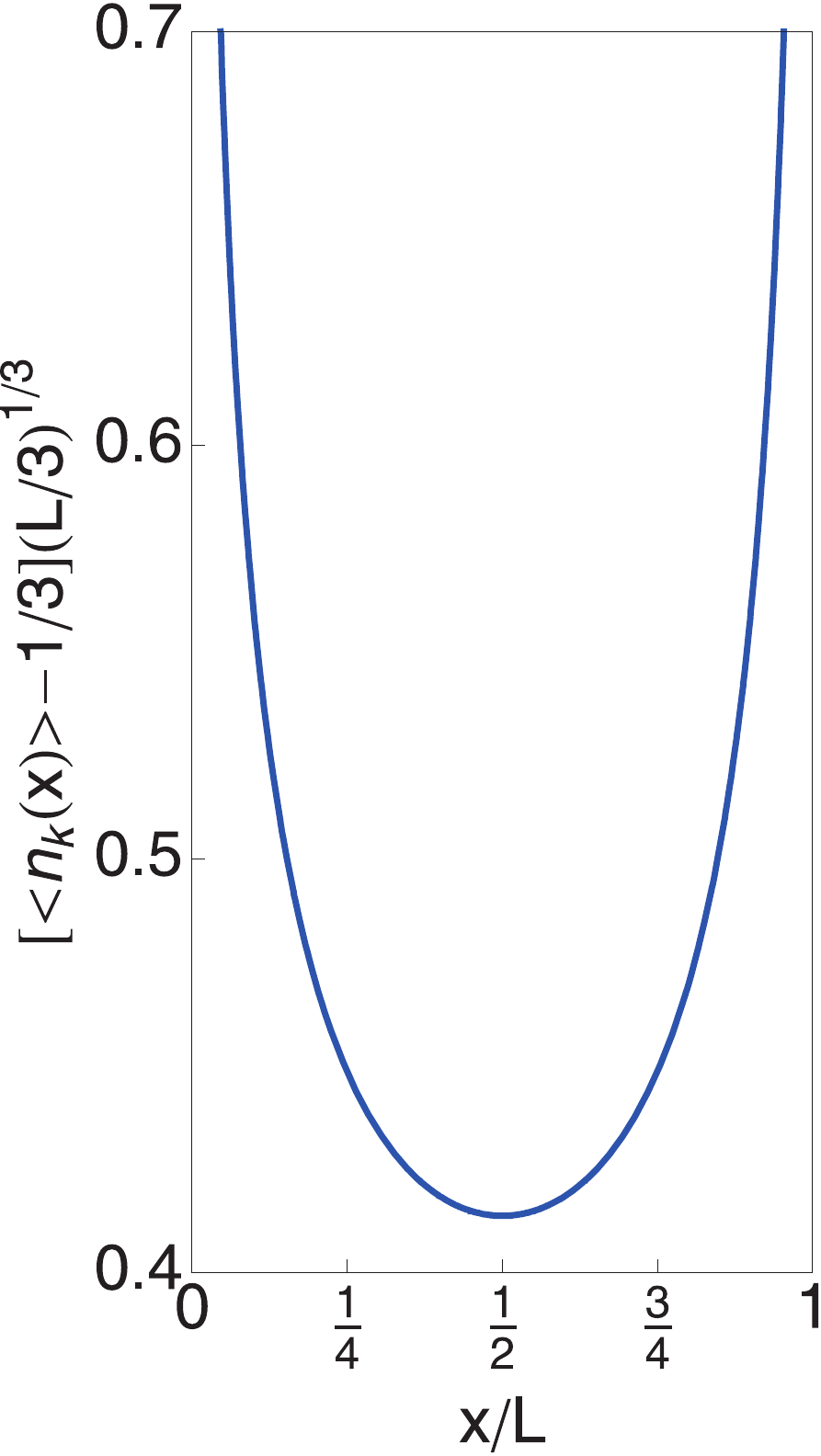}}
    \hspace{0.2cm}
     {\includegraphics[height=5.2cm]{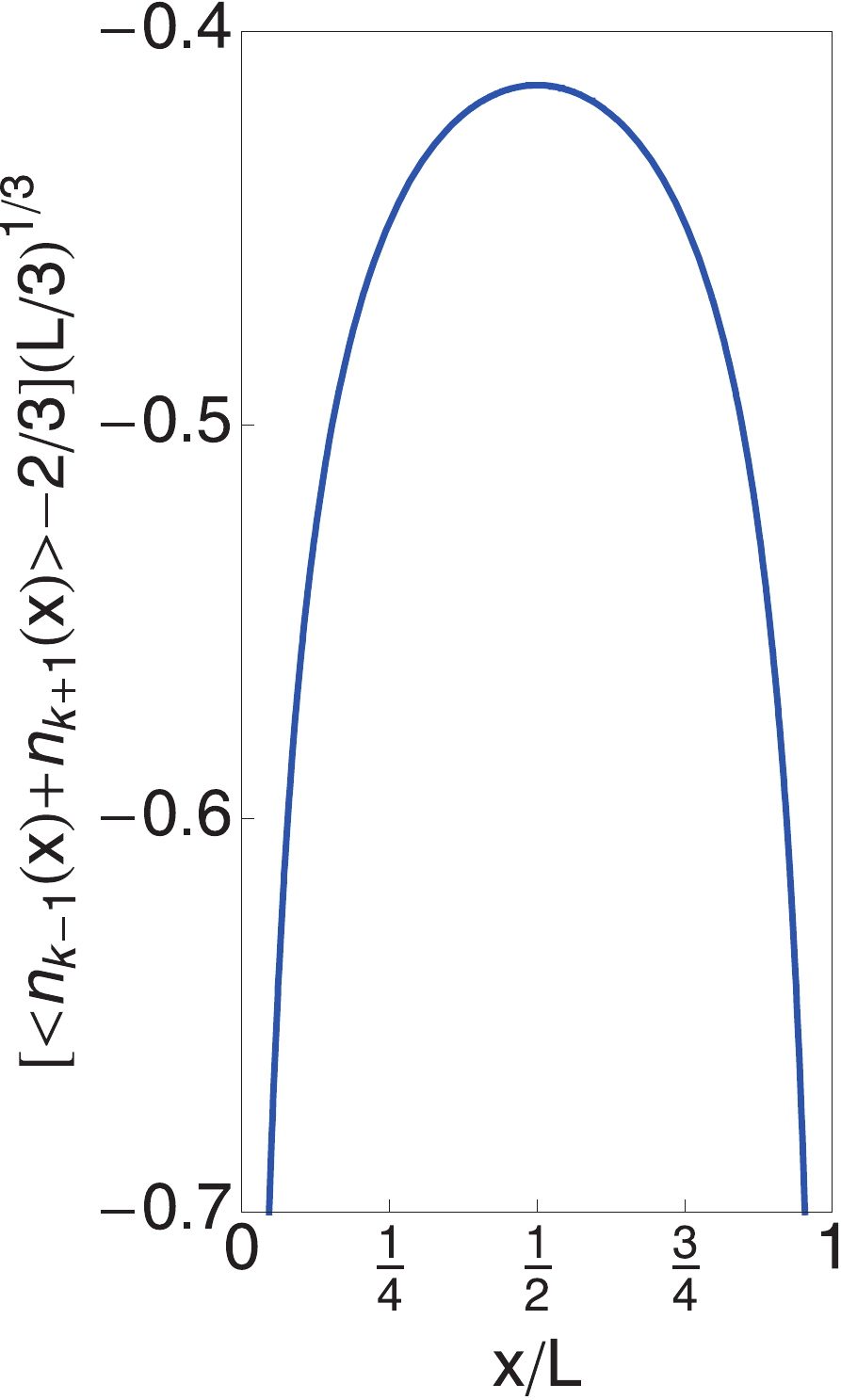}}}
     \caption{On the left we show the one-point functions for a chain of length $L_c=30$, this plot can be compared directly to the data presented in figure 1 in \cite{Beccaria05}. On the right we plot the finite size scaling functions, $(\langle n_k\rangle -1/3) (L/3)^{1/3}$ and $(\langle n_{k+1}+n_{k+2}\rangle -2/3) (L/3)^{1/3}$, to be compared with figure 2 in \cite{Beccaria05}.}
\end{figure}

Now remember that we identified the ground state of the chain of length $L=3j$ and open boundary conditions with the Ramond vacuum $V_{-1/2} \ket{0}$. To compute an expectation value we define the in-state as $\ket{R} \equiv \lim_{z \rightarrow 0} V_{-1/2} (z) \ket{0}$ and the out-state as $\bra{R} \equiv \lim_{z \rightarrow \infty} \bra{0} V_{1/2} (z) z^{1/12}$. These definitions ensure that
\begin{eqnarray}
 \langle R | R \rangle &=&\lim_{z_2 \rightarrow 0, z_1 \rightarrow \infty}\bra{0} V_{1/2} (z_1) z_1^{1/12} V_{-1/2} (z_2) \ket{0} \nonumber\\
 &=& \lim_{z_2 \rightarrow 0, z_1 \rightarrow \infty} z_1^{1/12} (z_1 -z_2)^{-1/12}\nonumber\\
 &=& 1.\nonumber
\end{eqnarray}
It thus follows that
\begin{eqnarray}
 \bra{R} V_{0,1} (x) \ket{R}_{\textrm{strip}} &=& \lim_{z_2 \rightarrow 0, z_1 \rightarrow \infty} \left( \frac{\pi \imath}{L}  \right)^{1/3}  (z \z)^{1/6} \bra{0} V_{1/2} (z_1) z_1^{1/12} V_{0,1} (z,\z) V_{-1/2} (z_2) \ket{0}_{\textrm{UHP}}  \nn
 &=& \lim_{z_2 \rightarrow 0, z_1 \rightarrow \infty} \left( \frac{\pi \imath}{L}  \right)^{1/3}  (z \z)^{1/6} \bra{0} V_{1/2} (z_1) z_1^{1/12} V_{1} (z) V_{-1} (z^*) V_{-1/2} (z_2) \ket{0}  \nn
&=& (-1)^{1/6}  \left( \frac{\pi \imath}{L} \right)^{1/3} \z^{1/3} (z - \z)^{-1/3}. \nonumber
\end{eqnarray}
In the last step we used (\ref{eq:correlator}). Equivalently, we find
\begin{eqnarray}
 \bra{R} V_{0,-1} (x) \ket{R}_{\textrm{strip}} &=& (-1)^{1/6} \left( \frac{\pi \imath}{L} \right)^{1/3} z^{1/3} (z - \z)^{-1/3}\nonumber.
\end{eqnarray}
Combining all the above, and setting $A_1=A_{-1}=A$, we obtain\footnote{To obtain these results, that are in good agreement with numerical observations, we write $(-1)^{1/6}=e^{\pm \pi \imath/6}$, where we use the $\pm$ sign for $V_{0,\pm1}$ respectively. This ensures that the results for $V_{0,\pm1}$ are related via complex conjugation.}
\begin{eqnarray}
 3\langle n_k\rangle  &=& 1 + 2A \left( \frac{\pi }{2 L} \right)^{1/3} \frac{\cos(\pi (x-L/2)/3L)}{\sin^{1/3}(\pi x/L)} \nonumber\\
 3\langle n_{k+1}\rangle  &=& 1 + 2A \left( \frac{\pi }{2 L} \right)^{1/3} \frac{\sin(\pi (x-L)/3L)}{\sin^{1/3}(\pi x/L)} \nonumber\\
 3\langle n_{k-1}\rangle  &=& 1 - 2A \left( \frac{\pi }{2 L} \right)^{1/3}  \frac{\sin (\pi x/3L)}{\sin^{1/3}(\pi x/L)}.
\end{eqnarray}
These equations clearly reproduce the observed scaling behavior. Comparison with the numerics suggests that we should choose $A \approx 0.77$. Finally, we argue that we should identify the width of the strip $L$ with $L_{c}+3$, where $L_c$ is the length of the chain. The way we understand this, is that the open chain can be obtained from a periodic chain with three sites extra, by pinning one particle to a certain site. Due to the hard-core character of the particles, the neighboring two sites must be empty. One thus effectively takes out three sites from the system and is left with an open chain. Finally, this implies that the sites -1 and $L_c+2$ are identified with the boundaries of the strip: $x=0$ and $x=L$. Consequently, an arbitrary site $p$ of the chain should be identified with the point $x=p+1$ on the strip. In figure \ref{fig:beccaria30} we show the one-point functions for a chain of length $L_c=30$, this plot can be compared directly to the data presented in figure 1 in \cite{Beccaria05}. In figure \ref{fig:beccariafss} we plot the finite size scaling functions to be compared with figure 2 in \cite{Beccaria05}. The agreement with the data of Beccaria et. al. is quite convincing, however, it seems to get poorer upon approaching the boundary of the system. In particular, the value for $\langle n_1\rangle $, which they obtain analytically using sophisticated number theoretical methods, cannot be reproduced from the field theory.

We can use the scaling functions to compute $F_i \equiv \sum_{k,k \mod 3=i} \langle n_k\rangle $, which is the average fermion number in the three branches distinguished by the site number modulo three. Replacing the sum by an integral and using that
\begin{eqnarray}
 \int_0^1 \frac{\cos(\pi (x-1/2)/3)}{\sin^{1/3}(\pi x)} d x &=& 2^{1/3}, \nonumber\\
\int_0^1 \frac{\sin (\pi x/3)}{\sin^{1/3}(\pi x)} d x &=& -2^{-2/3}, \nonumber
\end{eqnarray}
we find
\begin{eqnarray}
 F_2 &=& L/9 + A \frac{2}{3} \left( \frac{\pi }{2 L} \right)^{1/3}  \frac{L}{3} 2^{1/3} \approx F/3 +  0.52\ F^{2/3} , \nonumber\\
F_1=F_0 &=& L/9 - A \frac{2}{3} \left( \frac{\pi }{2 L} \right)^{1/3}  \frac{L}{3} 2^{-2/3} \approx F/3 - 0.26\ F^{2/3} ,\nonumber
\end{eqnarray}
where we used $ A \approx 0.77$ and $F=L/3$.
% The factors $L/3$ result from replacing the sum by an integral: $\sum_{k,k \mod 3=i} = L/3 \int_0^1 dx$. With $e^{\pi \imath /6} A_1 \approx 0.77$ and $F=L/3$ this gives
% \begin{eqnarray}
%  F_2 &\approx& F/3 +  0.52 F^{2/3} , \nonumber\\
% F_1=F_0 &\approx& F/3 - 0.26 F^{2/3} .\nonumber
% \end{eqnarray}
Consequently, the oscillation in occupation number with period three in the site number, does not lead to a difference in the average fermion number in the three branches. However, the subleading term goes as $F^{-1/3}$, which goes to zero only very slowly for $F \to \infty$.

On a qualitative level the $\mathbb{Z}_3$ substructure can probably be interpreted as follows. Remember that on the periodic chain we have two ground states, which are plane waves with opposite momenta differing by $2\pi/3$. It seems that for open boundary conditions these two ground states combine into a standing wave, which explains the observed density fluctuations.

\subsection{Closed boundary conditions}\label{sec:pbc1pntfn}
For the chain with periodic boundary conditions and length $L=0 \mod 3$ there are two ground states. The two ground states with momenta $\pm \pi/3+f_{\textrm{GS}} \pi \mod 2 \pi$ correspond to $ \ket{k_1}= \lim_{z \to 0} V_{0,1/2} (z) \ket{0}$ and $ \ket{k_2}= \lim_{z \to 0} V_{0,-1/2} (z) \ket{0}$ respectively. Using the same techniques as above we can compute the one-point functions using a conformal mapping from the cylinder to the plane. We find
\beq
\bra{k_i}V_{0,1} \ket{k_j} &=& \left\{ \begin{array}{ll} \left( \frac{2 \pi}{L} \right)^{1/3} & \textrm{for $i=1$ and $j=2$} \\ 0& \textrm{otherwise}   \end{array} \right. \nn
\bra{k_i}V_{0,-1} \ket{k_j} &=& \left\{ \begin{array}{ll} \left( \frac{2 \pi}{L} \right)^{1/3} & \textrm{for $i=2$ and $j=1$} \\ 0& \textrm{otherwise}   \end{array} \right. \nn
\bra{k_i}V_{0,0} \ket{k_j} &=& \left\{ \begin{array}{ll} 1& \textrm{for $i=j$} \\ 0& \textrm{otherwise}   \end{array} \right. \nonumber
\eeq
If we define the states $\ket{\pm}=(\ket{k_1}\pm \ket{k_2})/\sqrt{2}$, we obtain
\beq
\bra{\pm} n_k \ket{\pm} &=& \frac{1}{3} \pm \frac{1}{3} \left( \frac{2 \pi}{L} \right)^{1/3} A \nn
\bra{\pm} n_{k+1} \ket{\pm}=\bra{\pm} n_{k-1} \ket{\pm} &=& \frac{1}{3} \mp \frac{1}{6} \left( \frac{2 \pi}{L} \right)^{1/3} A \nonumber
\eeq
These expressions can be compared with the results of \cite{Fendley10b}, where they find $\bra{+} n_k \ket{+}-\bra{-} n_k \ket{-}= \left( \frac{3}{L}\right)^{1/3} \frac{4 \Gamma (2/3)}{3 \Gamma(1/3)}$. From this we extract
\beq
A= 2 \left( \frac{3}{2 \pi}\right)^{1/3} \frac{ \Gamma (2/3)}{\Gamma(1/3)} \approx 0.7901,
\eeq
which is in good agreement with the value ($A \approx 0.77$) we obtained in the previous section.

\section{Conclusions}
We have presented a full dictionary that relates the supersymmetric model for lattice fermions on the 1D chain to the $\mathcal{N}=(2,2)$ superconformal field theory with central charge $c=1$. As an example we have shown how this thorough understanding of the continuum limit can be employed to compute properties such as the particle density. It is remarkable that the continuum theory can reproduce the site dependent oscillations in the density that are observed for finite systems. An obvious extension of this work is to analyze two-point functions. Furthermore, this work could be extended to the generalized supersymmetric models that are labelled by an index $k$ \cite{fendley-2003-36}. The generalized supersymmetric models are found to have a quantum critical point described by the $k$-th superconformal minimal model. A similar dictionary relating the $k$-th supersymmetric model with the $k$-th superconformal minimal model could thus be developed. Finally, the study of the supersymmtric model on the chain away from the critical point \cite{Fendley10a, Fendley10b, Huijse11b} is likely to benefit from the detailed understanding of the model at the critical point presented here.

\section*{Acknowledgements}
The author would like to thank K. Schoutens for numerous valuable discussions, E. Verlinde for helpful discussions on the particle density and one-point function computations and P. Calabrese for discussions on the computation of the Fermi velocity. This work was carried out as part of the authors PhD research at the Institute for Theoretical Physics Amsterdam. We acknowledge financial support from the Netherlands Organisation for Scientific Research (NWO).

\bibliographystyle{unsrt}
\bibliography{literature}

\newpage

\appendix
\section{Results from spectral flow analysis}\label{app:spflow}

\begin{table}[h!]
 \caption{In this table we summarize the values we extract from the parabola fits for $E_{1/2}/c$ and $\tilde{Q}_{1/2}/c$ for various system sizes. Note that for each chain length we give two pairs of extracted values. They correspond to two different highest weight states in the Neveu-Schwarz sector. The values are extracted from a fit to the flow of that particular highest weight state from the Ramond to the NS sector. To be precise, for the chains with length $L=3j$ the middle two columns are extracted from a fit to the flow of the state corresponding to the field $V_{0,-1/2}$ to the state corresponding to the field $V_{0,0}$, whereas the last two columns are extracted from a fit to the flow corresponding to $V_{0,1/2} \to V_{0,1}$. For the chains with length $L=3j\pm1$ the middle two columns follow from $V_{\mp1/3,0} \to V_{\mp1/3,1/2}$, and the last two columns from $V_{\mp1/3,-1} \to V_{\mp1/3,-1/2}$. The theoretically predicted values  for $L=3j$ are $(E_{1/2}/c,\tilde{Q}_{1/2}/c)=(-1/12,0)$ and $(1/4,2/3)$ for the first and second column respectively. For $L=3j\pm1$ the theory predicts $(1/12,1/3)$ for the first and  $(1/12,-1/3)$ for the second column. \label{tab:spflowchain}}
\begin{center}
\begin{tabular}{|c|c|c|c|c|c|}
\hline
chain & fermion & & & & \\
length & number & $E_{1/2}/c$ & $\tilde{Q}_{1/2}/c$ & $E_{1/2}/c$ & $\tilde{Q}_{1/2}/c$ \\
\hline
 6 & 2 & -0.085 & -0.004 & 0.339 & 0.846 \\
 9 & 3 & -0.084 & -0.002 & 0.285 & 0.738 \\
 12 & 4 & -0.084 & -0.001 & 0.270 & 0.706 \\
 15 & 5 & -0.084 & -0.001 & 0.263 & 0.692 \\
 18 & 6 & -0.084 & 0.000 & 0.259 & 0.685 \\
 21 & 7 & -0.083 & 0.000 & 0.257 & 0.680 \\
 24 & 8 & -0.083 & 0.000 & 0.255 & 0.677 \\
 27 & 9 & -0.083 & 0.000 & 0.254 & 0.675\\
\hline
 5 & 2 & 0.086 & 0.338 & 0.138 & -0.560 \\
 8 & 3 & 0.084 & 0.335 & 0.101 & -0.407 \\
 11 & 4 & 0.084 & 0.334 & 0.093 & -0.372 \\
 14 & 5 & 0.084 & 0.334 & 0.089 & -0.357 \\
 17 & 6 & 0.084 & 0.334 & 0.087 & -0.350 \\
 20 & 7 & 0.083 & 0.334 & 0.086 & -0.346 \\
 23 & 8 & 0.083 & 0.334 & 0.086 & -0.343 \\
 26 & 9 & 0.083 & 0.334 & 0.085 & -0.341 \\
\hline
 7 & 2 & 0.085 & 0.336 & 0.108 & -0.436 \\
 10 & 3 & 0.084 & 0.335 & 0.095 & -0.381 \\
 13 & 4 & 0.084 & 0.334 & 0.09 & -0.362 \\
 16 & 5 & 0.084 & 0.334 & 0.088 & -0.352 \\
 19 & 6 & 0.084 & 0.334 & 0.087 & -0.347 \\
 22 & 7 & 0.083 & 0.334 & 0.086 & -0.344 \\
 25 & 8 & 0.083 & 0.334 & 0.085 & -0.342\\
\hline
\end{tabular}
\end{center}
\end{table}

\end{document}

%% file: SCFT.pdf_t
\begin{picture}(0,0)%
\includegraphics{SCFT.pdf}%
\end{picture}%
\setlength{\unitlength}{3947sp}%
\begingroup\makeatletter\ifx\SetFigFont\undefined%
\gdef\SetFigFont#1#2#3#4#5{%
  \reset@font\fontsize{#1}{#2pt}%
  \fontfamily{#3}\fontseries{#4}\fontshape{#5}%
  \selectfont}%
\fi\endgroup%
\begin{picture}(11805,5868)(811,-14323)
\put(7426,-13486){\makebox(0,0)[lb]{\smash{{\SetFigFont{20}{24.0}{\rmdefault}{\mddefault}{\updefault}{\color[rgb]{0,0,0}$m$}%
}}}}
\put(8701,-9436){\makebox(0,0)[lb]{\smash{{\SetFigFont{20}{24.0}{\rmdefault}{\mddefault}{\updefault}{\color[rgb]{0,0,0}$E$}%
}}}}
\put(12601,-12961){\makebox(0,0)[lb]{\smash{{\SetFigFont{20}{24.0}{\rmdefault}{\mddefault}{\updefault}{\color[rgb]{0,0,0}$n$}%
}}}}
\put(3976,-8686){\makebox(0,0)[lb]{\smash{{\SetFigFont{20}{24.0}{\rmdefault}{\mddefault}{\updefault}{\color[rgb]{0,0,0}$m$}%
}}}}
\put(826,-11761){\makebox(0,0)[lb]{\smash{{\SetFigFont{20}{24.0}{\rmdefault}{\mddefault}{\updefault}{\color[rgb]{0,0,0}$n$}%
}}}}
\put(5176,-10486){\makebox(0,0)[lb]{\smash{{\SetFigFont{20}{24.0}{\rmdefault}{\mddefault}{\updefault}{\color[rgb]{0,0,0}$G^+_R$}%
}}}}
\put(5176,-12886){\makebox(0,0)[lb]{\smash{{\SetFigFont{20}{24.0}{\rmdefault}{\mddefault}{\updefault}{\color[rgb]{0,0,0}$G^-_L$}%
}}}}
\put(2551,-12886){\makebox(0,0)[lb]{\smash{{\SetFigFont{20}{24.0}{\rmdefault}{\mddefault}{\updefault}{\color[rgb]{0,0,0}$G^+_R$}%
}}}}
\put(2626,-10486){\makebox(0,0)[lb]{\smash{{\SetFigFont{20}{24.0}{\rmdefault}{\mddefault}{\updefault}{\color[rgb]{0,0,0}$G^-_L$}%
}}}}
\end{picture}%

%% file: openchain1.pdf_t
\begin{picture}(0,0)%
\includegraphics{openchain1.pdf}%
\end{picture}%
\setlength{\unitlength}{3947sp}%
\begingroup\makeatletter\ifx\SetFigFont\undefined%
\gdef\SetFigFont#1#2#3#4#5{%
  \reset@font\fontsize{#1}{#2pt}%
  \fontfamily{#3}\fontseries{#4}\fontshape{#5}%
  \selectfont}%
\fi\endgroup%
\begin{picture}(9627,8272)(586,-9056)
\put(601,-1111){\makebox(0,0)[lb]{\smash{{\SetFigFont{25}{30.0}{\rmdefault}{\mddefault}{\updefault}{\color[rgb]{0,0,0}$E$}%
}}}}
\put(9901,-8461){\makebox(0,0)[lb]{\smash{{\SetFigFont{25}{30.0}{\rmdefault}{\mddefault}{\updefault}{\color[rgb]{0,0,0}$m$}%
}}}}
\put(9976,-8911){\makebox(0,0)[lb]{\smash{{\SetFigFont{25}{30.0}{\rmdefault}{\mddefault}{\updefault}{\color[rgb]{0,0,0}$\tilde{f}$}%
}}}}
\end{picture}%

%% file: 1Dchain.bbl
\begin{thebibliography}{10}

\bibitem{Huijse08b}
L.~Huijse, J.~Halverson, P.~Fendley, and K.~Schoutens.
\newblock Charge frustration and quantum criticality for strongly correlated
  fermions.
\newblock {\em Phys. Rev. Lett.}, 101:146406, 2008.

\bibitem{Beccaria05}
M.~Beccaria and G.~F.~De Angelis.
\newblock Exact ground state and finite size scaling in a supersymmetric
  lattice model.
\newblock {\em Phys. Rev. Lett.}, 94:100401, 2005.

\bibitem{fendley-2003-90}
P.~Fendley, K.~Schoutens, and J.~{de Boer}.
\newblock {Lattice models with $\mathcal{N}=2$ supersymmetry}.
\newblock {\em Phys. Rev. Lett.}, 90:120402, 2003.

\bibitem{fendley-2003-36}
P.~Fendley, B.~Nienhuis, and K.~Schoutens.
\newblock Lattice fermion models with supersymmetry.
\newblock {\em J. Phys. A}, 36:12399, 2003.

\bibitem{fendley-2005-95}
P.~Fendley and K.~Schoutens.
\newblock Exact results for strongly-correlated fermions in 2+1 dimensions.
\newblock {\em Phys. Rev. Lett.}, 95:046403, 2005.

\bibitem{Fendley05}
P.~Fendley, K.~Schoutens, and H.~van Eerten.
\newblock Hard squares at negative activity.
\newblock {\em J. Phys. A}, 38:315, 2005.

\bibitem{vanEerten05}
H.~van Eerten.
\newblock Extensive ground state entropy in supersymmetric lattice models.
\newblock {\em J. Math. Phys.}, 46:123302, 2005.

\bibitem{Jonsson06}
J.~Jonsson.
\newblock Hard squares with negative activity and rhombus tilings of the plane.
\newblock {\em Electr. J. Comb.}, 13(1):\#R67, 2006.

\bibitem{Jonsson05p}
J.~Jonsson.
\newblock Certain homology cycles of the independence complex of grid graphs.
\newblock {\em Discrete Comput. Geom.}, 2010.

\bibitem{Huijse10}
L.~Huijse and K.~Schoutens.
\newblock Supersymmetry, lattice fermions, independence complexes and
  cohomology theory.
\newblock {\em Adv. Theor. Math. Phys.}, 14.2, 2010.
\newblock Preprint [ArXiv:0903.0784].

\bibitem{Huijse10b}
L.~Huijse and K.~Schoutens.
\newblock Quantum phases of supersymmetric lattice models.
\newblock In P.~Exner, editor, {\em {XVITH} {International} CONGRESS ON
  MATHEMATICAL PHYSICS}, pages 635--639. World Scientific, 2010.
\newblock Preprint [ArXiv:0910.2386].

\bibitem{Fendley10a}
P.~Fendley and C.~Hagendorf.
\newblock Exact and simple results for the xyz and strongly interacting fermion
  chains.
\newblock {\em J. Phys. A}, 43:402004, 2010.

\bibitem{Fendley10b}
P.~Fendley and C.~Hagendorf.
\newblock Ground-state properties of a supersymmetric fermion chain.
\newblock {\em J. Stat. Mech.}, 02:P02014, 2011.

\bibitem{Huijse08a}
L.~Huijse and K.~Schoutens.
\newblock Superfrustration of charge degrees of freedom.
\newblock {\em EPJ B}, 64:543--550, 2008.

\bibitem{HuijseT10}
L.~Huijse.
\newblock A supersymmetric model for lattice fermions.
\newblock PhD Thesis, 2010.

\bibitem{Witten82}
E.~Witten.
\newblock Constraints on supersymmetry breaking.
\newblock {\em Nucl. Phys. B}, 202(2):253--316, 1982.

\bibitem{Thacker81}
H.~B. Thacker.
\newblock Exact integrability in quantum field theory and statistical systems.
\newblock {\em Rev. Mod. Phys.}, 53(2):253--285, 1981.

\bibitem{Friedan88}
D.~Friedan and A.~Kent.
\newblock Supersymmetric critical phenomena and the two dimensional gaussian
  model.
\newblock In Conformal Invariance and Applications to Statistical Mechanics,
  eds. C. Itzykson, H. Saleur, and J.B. Zuber (World Scientific, Singapore,
  1988), pp. 578--579, 1988.

\bibitem{Waterson86}
G.~{Waterson}.
\newblock Bosonic construction of an $\mathcal{N}=2$ extended superconformal
  theory in two dimensions.
\newblock {\em Phys. Lett. B}, 171:77--80, 1986.

\bibitem{Affleck88}
I.~Affleck.
\newblock Field theory methods and quantum critical phenomena.
\newblock In E.~Brezin and J.~Zinn-Justin, editors, {\em Fields, strings,
  critical phenomena: proceedings of Les Houches Summer School in Theoretical
  Physics 1988.}, volume Session 49. North-Holland, 1990.

\bibitem{Boucher86}
W.~Boucher, D.~Friedan, and A.~Kent.
\newblock Determinant formulae and unitarity for the n = 2 superconformal
  algebras in two dimensions or exact results on string compactification.
\newblock {\em Phys. Lett. B}, 172(3-4):316--322, 1986.

\bibitem{DiVecchia85}
P.~{Di Vecchia}, J.~L. Petersen, and H.~B. Zheng.
\newblock $\mathcal{N}=2$ extended superconformal theories in two dimensions.
\newblock {\em Phys. Lett. B}, 162(4-6):327--332, 1985.

\bibitem{Feigin82}
B.~L. Feigin and D.~B. Fuchs.
\newblock Skew-symmetric differential operators on the line and {Verma} modules
  over the {Virasoro} algebra.
\newblock {\em Functs. Anal. Prilozhen}, 16:47, 1982.

\bibitem{Feigin85}
B.~L. Feigin and D.~B. Fuchs.
\newblock {Verma} modules over the {Virasoro} algebra.
\newblock In L.~D. Faddeev and A.~A. Malcev, editors, {\em Topology,
  Proceedings of {Leningrad} conference, 1982, Lecture Notes in Mathematics},
  volume 1060. Springer, New York, 1985.

\bibitem{Blote86}
H.~W.~J. Bl\"ote, J.~L. Cardy, and M.~P. Nightingale.
\newblock Conformal invariance, the central charge, and universal finite-size
  amplitudes at criticality.
\newblock {\em Phys. Rev. Lett.}, 56(7):742--745, 1986.

\bibitem{Affleck86}
I.~Affleck.
\newblock Universal term in the free energy at a critical point and the
  conformal anomaly.
\newblock {\em Phys. Rev. Lett.}, 56(7):746--748, 1986.

\bibitem{Yu92}
N.~Yu and M.~Fowler.
\newblock Twisted boundary conditions and the adiabatic ground state for the
  attractive {XXZ Luttinger} liquid.
\newblock {\em Phys. Rev. B}, 46(22):14583--14593, 1992.

\bibitem{Schwimmer87}
A.~Schwimmer and N.~Seiberg.
\newblock Comments on the $\mathcal{N} = 2,3,4$ superconformal algebras in two
  dimensions.
\newblock {\em Phys. Lett. B}, 184(2-3):191--196, 1987.

\bibitem{Holzhey94}
C.~Holzhey, F.~Larsen, and F.~Wilczek.
\newblock Geometric and renormalized entropy in conformal field theory.
\newblock {\em Nucl. Phys. B}, 424(3):443--467, 1994.

\bibitem{Vidal03}
G.~Vidal, J.~I. Latorre, E.~Rico, and A.~Kitaev.
\newblock Entanglement in quantum critical phenomena.
\newblock {\em Phys. Rev. Lett.}, 90(22):227902, 2003.

\bibitem{Calabrese04}
P.~Calabrese and J.~Cardy.
\newblock Entanglement entropy and quantum field theory.
\newblock {\em J. Stat. Mech.}, 2004(06):P06002, 2004.

\bibitem{Korepin04}
V.~E. Korepin.
\newblock Universality of entropy scaling in one dimensional gapless models.
\newblock {\em Phys. Rev. Lett.}, 92(9):096402, 2004.

\bibitem{Campostrini}
M.~Campostrini.
\newblock Private communication.

\bibitem{Laflorencie06}
N.~Laflorencie, E.~S. S\o{}rensen, M.-S. Chang, and I.~Affleck.
\newblock Boundary effects in the critical scaling of entanglement entropy in
  {1D} systems.
\newblock {\em Phys. Rev. Lett.}, 96(10):100603, Mar 2006.

\bibitem{Cardy10}
J.~Cardy and P.~Calabrese.
\newblock Unusual corrections to scaling in entanglement entropy.
\newblock {\em J. Stat. Mech.}, 2010(04):P04023, 2010.

\bibitem{Calabrese10}
P.~Calabrese, J.~Cardy, and E.~Tonni.
\newblock Entanglement entropy of two disjoint intervals in conformal field
  theory {II}.
\newblock {\em J. Stat. Mech.}, 2011(01):P01021, 2011.

\bibitem{Huijse11b}
L.~Huijse, N.~Moran, J.~Vala, and K.~Schoutens.
\newblock In preparation, 2011.

\end{thebibliography}
